\documentclass[12pt,preprint ]{aastex}

\slugcomment{Draft: \today}

\shorttitle{New SPBe stars from MOST}

\begin{document}


\title{{\it MOST$^{\bigstar}$} detects SPBe pulsations in HD 127756 \& HD 217543:\\
Asteroseismic rotation rates independent of vsini}

\altaffiltext{$\bigstar$}{Based on data from the {\it MOST} satellite, a Canadian Space Agency mission, jointly operated by Dynacon Inc., the University of Toronto Institute of Aerospace Studies and the University of British Columbia with the assistance of the University of Vienna.} 
\author{ 
C. Cameron\altaffilmark{1}, 
H. Saio\altaffilmark{2},
R. Kuschnig\altaffilmark{3},
G.A.H. Walker\altaffilmark{4}, 
J.M. Matthews\altaffilmark{1},  
D.B. Guenther\altaffilmark{5}, 
A.F.J. Moffat\altaffilmark{6}, 
S.M. Rucinski\altaffilmark{7}, 
D. Sasselov\altaffilmark{8}, 
W.W. Weiss\altaffilmark{3}
} 

\altaffiltext{1}{Dept. of Physics and Astronomy, University of British Columbia, 
6224 Agricultural Road, Vancouver, BC V6T 1Z1, Canada; ccameron@phas.ubc.ca, 
matthews@phas.ubc.ca }

\altaffiltext{2}{Astronomical Institute, Graduate School of Science, Tohoku University, Sendai, 980-8578, Japan;
saio@astr.tohuku.ac.jp} 

\altaffiltext{3}{Institut f\"ur Astronomie, Universit\"at Wien 
T\"urkenschanzstrasse 17, A--1180 Wien, Austria;
a0206892@unet.univie.ac.at, rainer.kuschnig@univie.ac.at; weiss@astro.univie.ac.at}

\altaffiltext{4}{1234 Hewlett Place, Victoria, BC V8S 4P7, Canada;
gordonwa@uvic.ca}

\altaffiltext{5}{Department of Astronomy and Physics, St. Mary's University
Halifax, NS B3H 3C3, Canada;
guenther@ap.stmarys.ca}

\altaffiltext{6}{D\'ept. de physique, Univ. de Montr\'eal 
C.P.\ 6128, Succ.\ Centre-Ville, Montr\'eal, QC H3C 3J7, Canada;
and Obs. du mont M\'egantic;
moffat@astro.umontreal.ca}

\altaffiltext{7}{Dept. of Astronomy \& Astrophysics, David Dunlap Obs., Univ. Toronto 
P.O.~Box 360, Richmond Hill, ON L4C 4Y6, Canada;
rucinski@astro.utoronto.ca}

\altaffiltext{8}{Harvard-Smithsonian Center for Astrophysics, 
60 Garden Street, Cambridge, MA 02138, USA;
sasselov@cfa.harvard.edu}

\begin{abstract}
The \textit{MOST} (Microvariability and Oscillations of Stars) satellite has discovered SPBe (Slowly Pulsating Be) oscillations in the stars HD 127756 (B1/B2 Vne) and HD 217543 (B3 Vpe). For HD 127756, 30 significant frequencies are identified from 31 days of nearly continuous photometry; for HD 217543, up to 40 significant frequencies from 26 days of data. In both cases, the oscillations fall into three distinct frequency ranges, consistent with models of the stars. The variations are caused by nonradial g-modes (and possibly r-modes) distorted by rapid rotation and excited by the opacity mechanism near the iron opacity bump. A comparison of pulsation models and observed frequency groups yields a rotation frequency for each star, independently of vsini. The rotation rates of these stars, as well as those of the SPBe stars previously discovered by \textit{MOST},  HD 163868 and $\beta$ CMi, are all close to their critical values.
\end{abstract}


\section{Introduction}

Be stars are rapidly rotating B-type stars close to the main sequence that show or have shown emission lines in their photospheric spectra \citep[see][for a recent review]{por03}.  Some Be stars (especially those of early type) also exhibit line-profile variations indicating the presence of nonradial pulsations \citep[e.g.][]{riv03}. The \textit{ MOST} (Microvariability and Oscillations of STars) satellite \citep{Matthews04} photometrically detected multiple periods in three Be stars: $\zeta$ Oph \citep[O9.5 V;][]{wal05a}, HD 163868 [B1.5-5 Ve; see section \ref{hd163868mods} for details.]\citep{wal05b}, and $\beta$ CMi \citep[B8 Ve;][]{sai07}. \cite{wal05a} suggest the pulsations of $\zeta$ Oph ($\lesssim$ 20 c d$^{-1}$) are well modelled as a combination of low-order, radial and nonradial, p and g-modes; modified by rotation, and consistent with the $\beta$ Cephei-type variables. The oscillations in the latter two stars are attributed to high-order, nonradial g-mode pulsations excited by the $\kappa$ - mechanism near the Fe opacity bump ($\log T\approx5.3$),  as in the case of the slowly rotating SPB (slowly pulsating B) stars. Since the frequencies of high-order g-modes in the co-rotating frame of a rapidly rotating star are smaller than the rotation frequency $\Omega$, the pulsation frequencies in the observers' frame are grouped near, and are separated from other groups by, $\approx |m|\Omega$ according to the azimuthal order $m$. (Those frequency groups occur around $1.6$ c d$^{-1}$ and $3.3$ c d$^{-1}$ for HD 163868, and around $3.3$ c d$^{-1}$ for $\beta$ CMi.) This type of grouping characterizes the amplitude spectra of SPBe stars, and makes the periods of their light variations close to their rotation period, or half of it, just as in the $\lambda$ Eri variables \citep{bal95}.

Since the discovery of the SPBe variability in the aforementioned stars by the authors there have been observations of Be stars by \citet{uyt07} and \citet{gut07} that show seemingly similar characteristics. In particular, \citeauthor{uyt07} detected three periods (2.234 c d$^{-1}$, 4.713 c d$^{-1}$, 4.671 c d$^{-1}$) from their ground-based photometry in the Be star V2104 Cyg \citep[B5-7; as described in][]{uyt07} and \citeauthor{gut07} detected multiperiodic photometric variations in the two early type Be stars NW Ser (B2.5 IIIe) and V1446 Aql (B2 IVe).  Both space and ground-based campaigns of Be stars are yielding data that can be used in conjunction with models to determine the rotation periods of rapidly rotating stars asteroseismically from the observed frequency groupings alone.

In this paper, we report the {\it MOST} detections and modelling of SPBe pulsations in another two Be stars: HD 127756 and HD 217543. HD 127756 is a southern, early-type Be star (B1/B2 Vne; V=7.56 mag; $\delta =-61^\circ$) for which no $v\sin i$ value is available. HD 217543 (= V378 and = HR 8758) is an intermediate-type Be star with shell characteristics (B3 Vpe; V = 6.555 mag; $\delta = + 38^\circ42$).  \cite{2002ApJ...573..359A} report a value for $v\sin i$ of 305 km s$^{-1}$ for this star and \cite{1970CoAsi.239....1B} suggest a larger value ($370$ km s$^{-1}$). HD 217543 also shows marked variations in emission-line strength \citep{cop63}.  In addition, we present an alternate model of the SPBe star HD 163868 \citep{wal05b} and discuss the implications of the models and observations of all published data on the SPBe stars observed by \textit{MOST} to date.

\section{The {\it MOST} photometry and frequency analysis}\label{secanalysis}

The {\it MOST} satellite (launched on 30 June 2003) houses a 15/17.3 cm Rumak-Maksutov telescope feeding a CCD photometer through a single custom broadband optical filter (350 -- 700 nm); see \cite{MOST} for details. {\it MOST} observed HD 127756 and HD 217543 as guide stars for other Primary Science Targets in different observing runs. The guide stars are sampled by subrasters on the CCD and the photometry is primarily processed on-board before downlinking to Earth, by subtracting a mean sky value within the subraster after applying a specified threshold. Individual exposure times are set by the Attitude Control System (ACS) star-tracking requirements (about 0.5 and 1.5 sec for the two stars). Those exposures are ``stacked" on-board to build up signal-to-noise (S/N) and stacked samples are obtained roughly every 20 seconds. Table~\ref{obssumtab} summarises the observations. The target fields are outside the {\it MOST}'s Continuous Viewing Zone so there is a gap during part of each 101.4 min satellite orbit. The duty cycle during each orbit was 30.7\% for HD 127756 and 26.1\% for HD 217543,  but as can be seen in the light curves of Figures \ref{lc_hd127756} and \ref{lc_hd217543} below, the effective duty cycle for sampling the timescales of variability in these stars is close to 100\%.

The frequency analysis of the light curves was done using the CAPER code (\citealt{cam06}; also see \citealt{wal05b} and \citealt{sai06}). CAPER calculates a discrete Fourier transform of a time series and uses the position of the largest amplitude in the spectrum as an initial guess for the frequency, amplitude and phase parameters in a nonlinear least squares fit to the variability. A sinusoid is fitted to the data using all identified parameters and then subtracted from the original light curve. This process is repeated until a predefined S/N is reached in the amplitude spectrum.  

A peak with a S/N of $\sim 3.5$ is consistent with a detection $\gtrsim$ 2.5$\sigma$ \citep{kus97} and is adopted as our lower limit to the significance of the extracted periodicities. The S/N of each identified periodicity is estimated (before prewhitening that component) by taking the mean amplitude in a box around the identified peak and sigma clipping points until the mean converges. This is done to ensure that high amplitude peaks near the identified frequency do not skew the local mean amplitude. The S/N calculation method contains two potential sources of uncertainty: 1. The width of the box used to average the amplitude spectrum (the noise) and 2. Uncertainties in the fitted amplitude. We estimate the uncertainty in the noise calculation by varying the width of the averaging box from $\pm$0.5 c d$^{-1}$ to $\pm$5 c d$^{-1}$ in steps of 0.1 c d$^{-1}$ and then calculate an average noise and the standard error on that average noise. Once amplitude uncertainties are assessed from a bootstrap analysis (described below), we combine both to arrive at the final uncertainty in S/N. This uncertainty is dominated by the precision of  the amplitude parameter so to limit the size of our data tables we only report the amplitude uncertainties, but show the full errorbars in all S/N plots presented in the paper.   

Special care is taken to assess the precision of our fitted parameters and to identify frequencies that are possibly unresolved. Recently, \cite{breg07} discussed the difference between frequency resolution and the precision of fit parameters to time series data. Traditionally one estimates resolution in Fourier space as $T^{-1}$(Rayleigh criterion), where $T$ is the length of the observing run. \cite{loum78} suggest an upper limit of 1.5 $T^{-1}$ (roughly corresponding to the spacing between the main peak of the window function and the peak of its first sidelobe) be used when identifying periodicities directly from an amplitude spectrum. However, lower, and arguably more realistic,  estimates are used by \cite{kurtz}, who estimates frequency resolution as $0.5~T^{-1}$ (approximately the half-width of a peak in the amplitude spectrum), and by \cite{kallinger},  who suggest  $\sim0.25~T^{-1}$ can be used based on a large number of simulated data sets. Ultimately, the resolution of frequencies in Fourier space is a function of S/N (or significance) of each individual peak; see, for example, \cite{kallinger}, and the above criteria are only estimates used when determining the frequency resolution over the entire frequency range of interest. The precision of fitted parameters, on the other hand, can be estimated by refitting identified parameters to a large number of data sets created by sampling the fitted function in the same way as the data and adding random, normally distributed noise. This Monte Carlo procedure is used, for example, in Period04 \citep{period}. 

We assess the precision of our fit parameters and estimate our resolution using a type of bootstrap analysis \citep{cam06}. By randomly sampling the light curves of HD 127756 and HD 217543 (with the possibility of replacement) 100000 times and then refitting our parameters to those resampled data sets, we build distributions for each of the fit parameters, e.g. as in Fig.~\ref{distros_hd127756} discussed in the next section. We estimate the 1 and 3$\sigma$ uncertainties for each parameter as the width of the region, centred on the parameter in question, that contains 68 and 99\% of the bootstrap realisations, respectively. Notice that this differs from a Monte Carlo procedure (as described above) in that there are no assumtions made about the noise of our resampled data sets (we only use the data) and that the frequency resolution of our data sets can be estimated because the windowing (determined from the temporal sampling of the data) of each of the resampled data sets is randomly changed. Thus, we test the robustness of our fit against both the inherent noise of the data and the sampling of the data as well.  

\subsection{HD 127756}\label{secHD127756}

\begin{figure}[hp]
\epsscale{.70}
\plotone{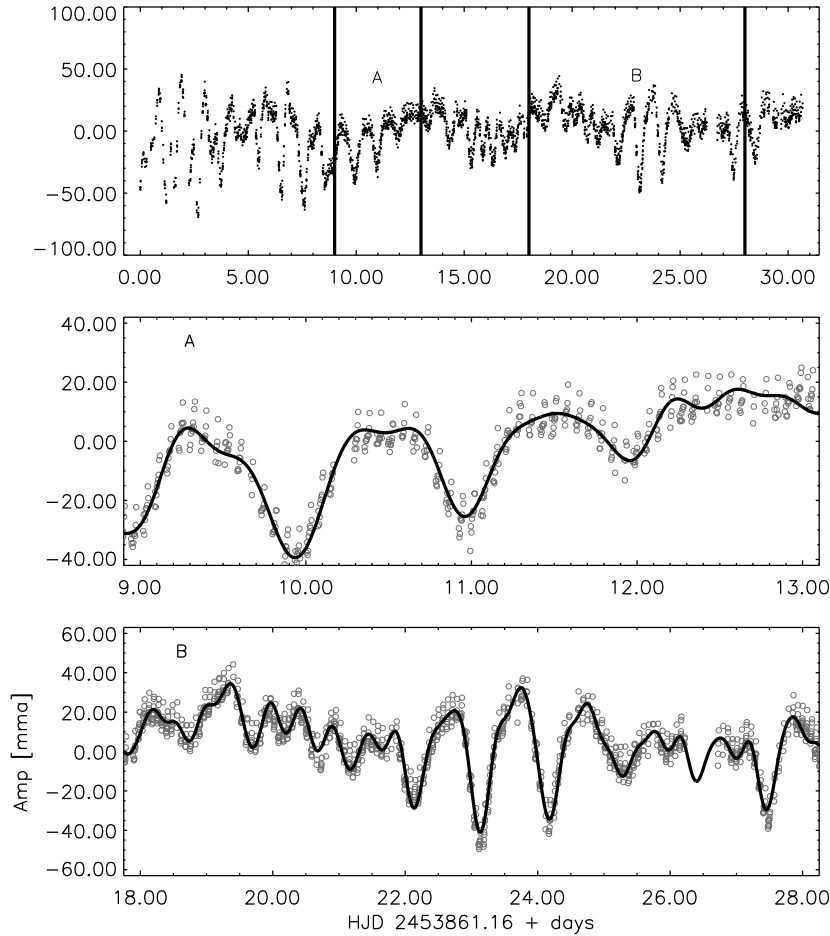}
\caption{{\it MOST} light curve of HD 127756. The top panel shows the entire light curve spanning a total of 30.7 days. The middle and the bottom panels are expanded light curves for the portions A and B, respectively, indicated in the top panel. Solid lines indicate the fit of the 30 significant frequencies (Table \ref{tab_hd127756}) from the frequency analysis of the full light curve. The short-term variability seen in the middle panel is a consequence of stray Earthshine modulated with the {\it MOST} satellite orbital period of $\sim 101.4$ min.
\label{lc_hd127756}}
\end{figure}

HD 127756 was observed for a total of 30.7 days by {\it MOST}. The light curve is shown in Fig.~\ref{lc_hd127756}, 
which shows clear variations with periods near $1$ day and $ 0.5$ day.(Note that {\it MOST} does not suffer from cycle/day aliasing due to daily gaps as experienced in single-site ground-based observations; these periodicities are intrinsic to the star.) Table~\ref{tab_hd127756} lists the frequencies, amplitudes, phases (referenced to the time of the first observation), the 1 and 3$\sigma$ uncertainties from the bootstrap analysis, and the S/N of the 30 most significant periodicities. The fit is shown superimposed over two zoomed sections of the light curve labeled A and B in the lower panels of Fig.~\ref{lc_hd127756}. 

\begin{figure}[htbp]
\epsscale{.75}
\plotone{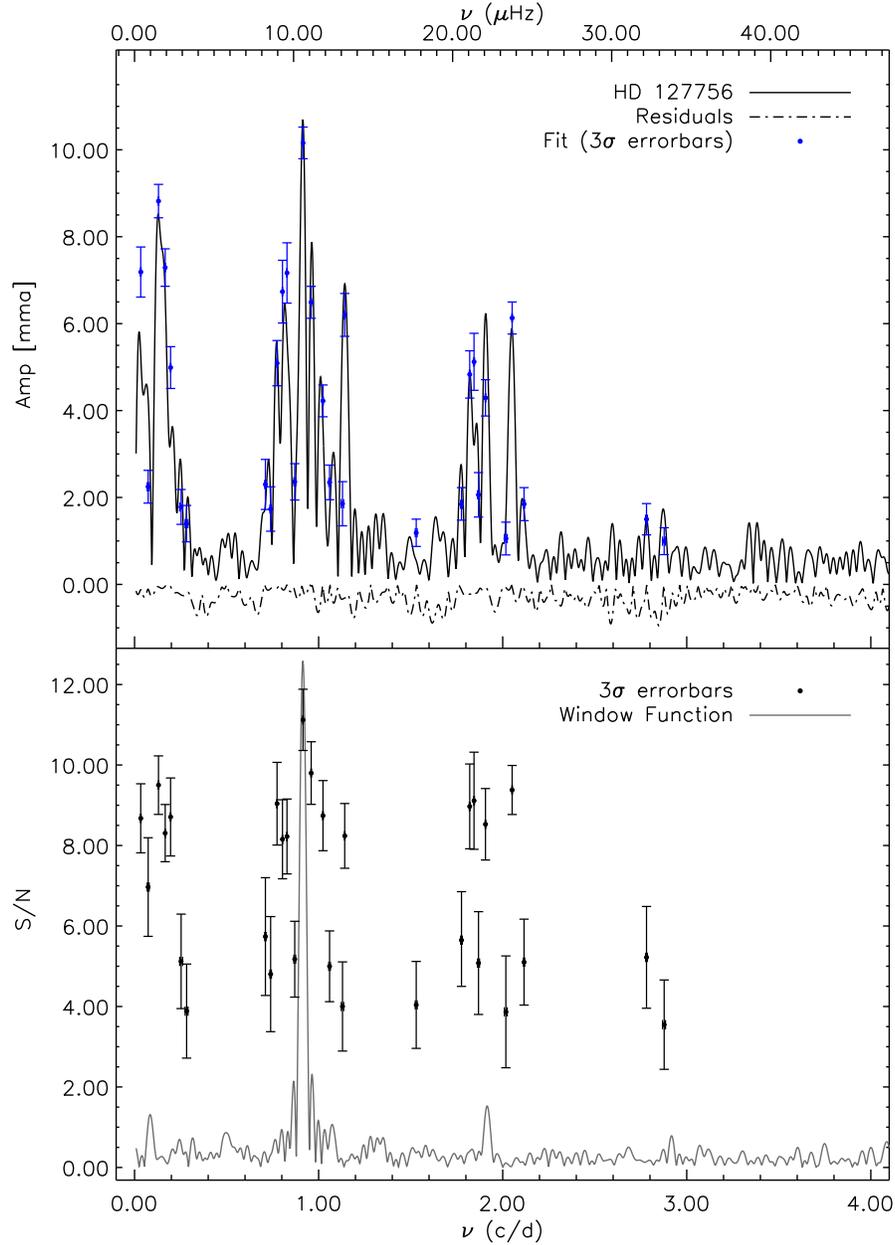}
\caption{(Top Panel) The Fourier amplitude spectrum of the light curve of HD 127756. Filled (blue) circles with $3\sigma$ error bars are the fitted parameters (see Table \ref{tab_hd127756}). The inverted dash - dot line is the residual amplitude spectrum obtained after the fit was subtracted from the light curve. (Lower Panel) The S/N of the identified periodicities with  $3\sigma$ uncertainties estimated from both the fitted amplitudes and frequencies and the mean of the amplitude spectrum (see section \ref{secanalysis} for details). The light grey line represents the window function of the data centred on the frequency with the largest amplitude and scaled to the maximum S/N for clarity.\label{ft_hd127756}}
\end{figure}

The amplitude spectrum of the data along with the fitted points and the residuals from the fit are shown in the top panel of Fig.~\ref{ft_hd127756}. Most of the frequencies gather into three groups; $\sim 0$ c d$^{-1}$, $\sim 1$ c d$^{-1}$, and $\sim 2$ c d$^{-1}$. This property is similar to the frequency groupings of the SPBe star HD 163868 \citep{wal05b}. The lower panel of Fig.~\ref{ft_hd127756} plots the S/N for each of the identified periodicities and the window function of the data. Among the frequencies listed in Table \ref{tab_hd127756}, $\nu{_1}=0.0335$ c~d$^{-1}$ and $\nu{_2}=0.0739$ c~d$^{-1}$ have the fewest observed cycles (close to the length of the run at $1/30.7 = 0.0326$ c~d$^{-1}$) and are included to reduce the scatter in the residuals from our fit. They may not be genuine stellar oscillation frequencies but it should be noted that we have not observed artifacts associated with the baselines of other {\it MOST} observations, especially with such a relatively large amplitude
of 7 mmag as in the case of $\nu{_1}$ here.

\begin{figure}[htbp]
\epsscale{.7}
\plotone{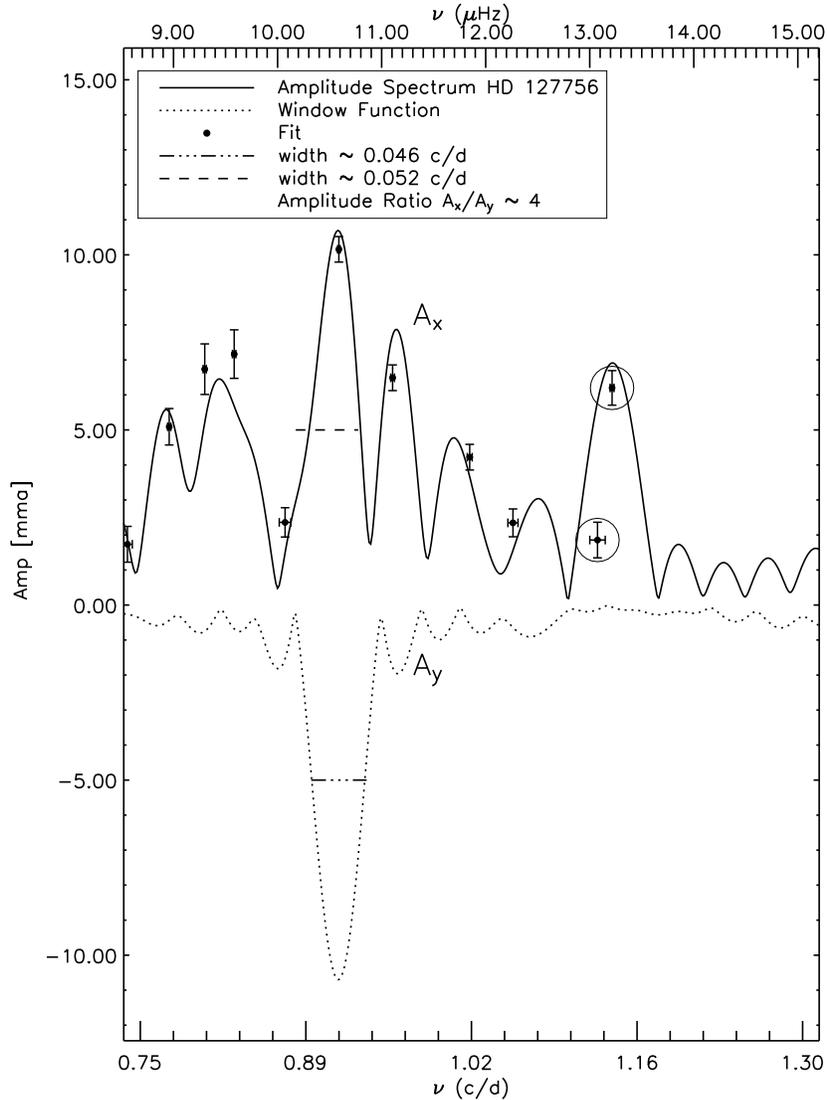}
\caption{A zoomed region around the largest peak in the amplitude spectrum of HD 127756. The window function is shown as the inverted, dotted line and the fit is shown as points with $3 \sigma$ errorbars. The asymmetry of the largest peak in the amplitude spectrum (width $\sim 0.052$ c d$^{-1}$ [ dashed line]) is compared to the width of the window function ($\sim 0.048$ c d$^{-1}$ [ dash dot line]). The amplitude of the first sidelobe of the window function (labelled A$_{\textrm{y}}$) is $\sim 4$ times smaller than the second largest peak in the amplitude spectrum at A$_{\textrm{x}}$. The resolution of frequencies $\nu_{18}$ and $\nu_{19}$ (both circled) is discussed in section \ref{secHD127756}. \label{ft_hd127756winft}} 
\end{figure}

\begin{figure}[htbp]
\epsscale{1.0}
\plottwo{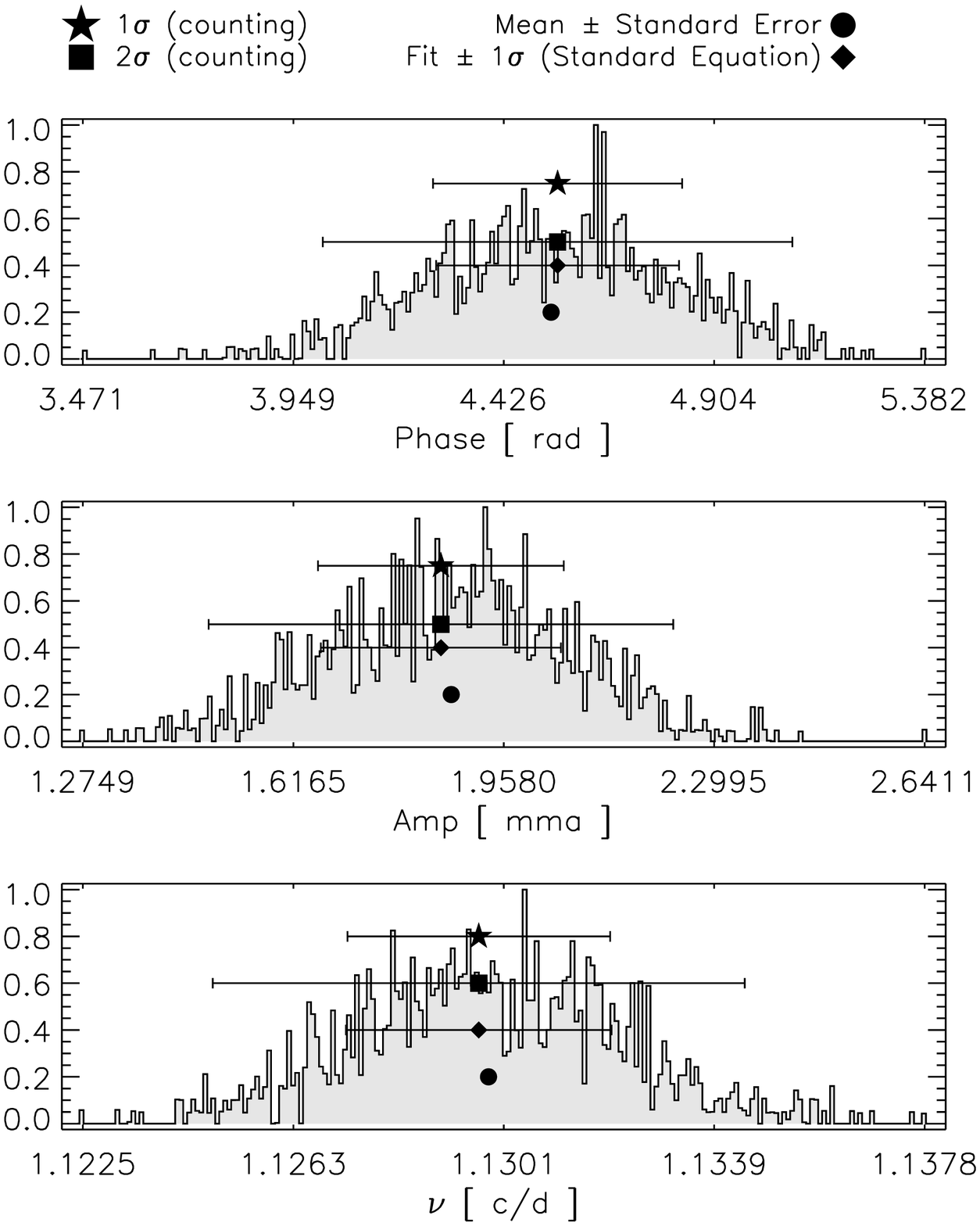}{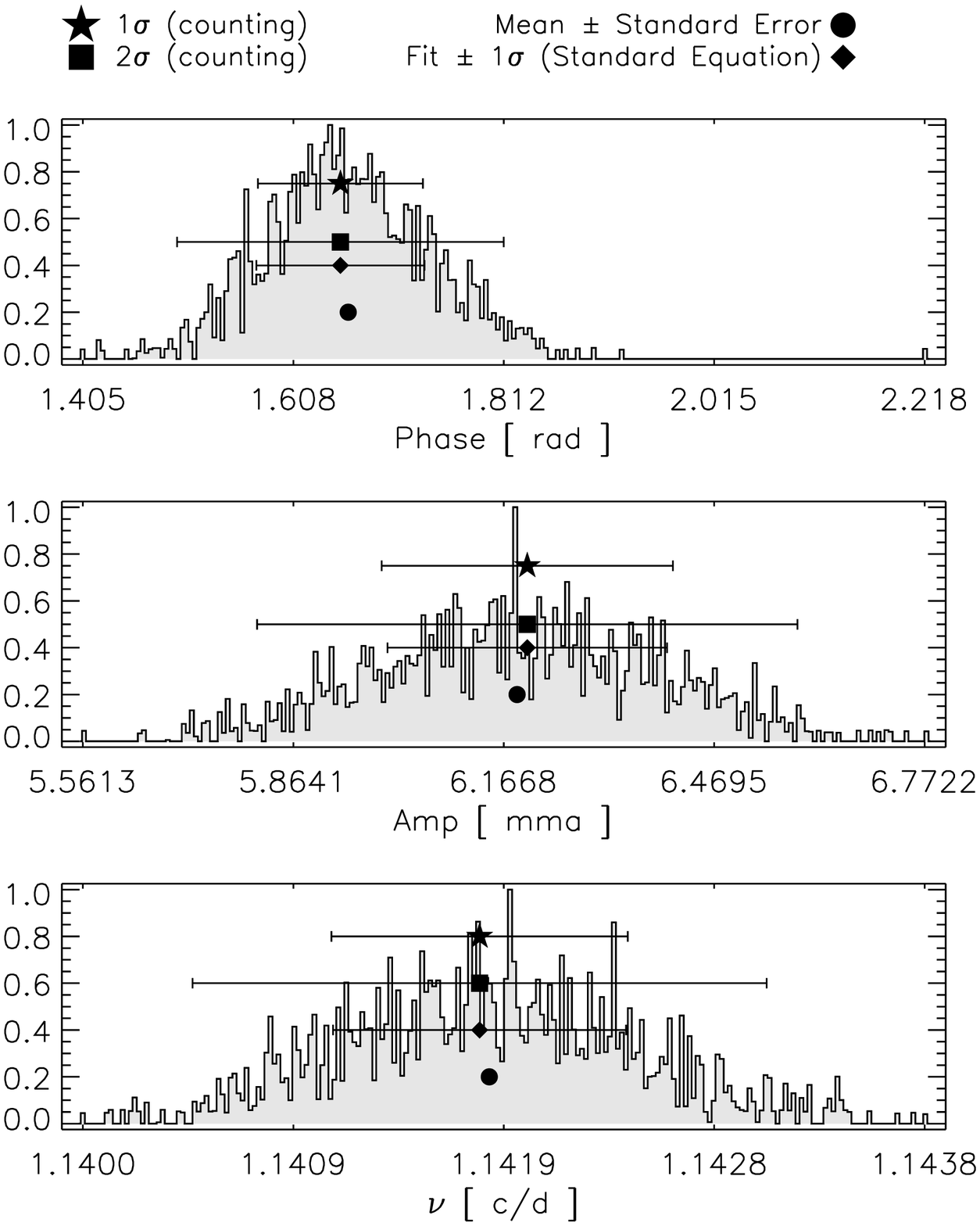}
\caption{A comparison of bootstrap distributions for parameter sets ($\nu_{18}$, $\textrm{A}_{18}$, $\phi_{18}$) and ($\nu_{19}$, $\textrm{A}_{19}$, $\phi_{19}$)  [see Table \ref{tab_hd127756}] for 100000 realisations of the HD 127756 light curve. The top panels are distributions for the fitted phase ($\phi$) while the middle and lower panels show distributions for the amplitude (A) and frequency ($\nu$) parameters, respectively. In each panel symbols are shown (from top to bottom) for the $1 \sigma~(\bigstar)$ and $2 \sigma~(\blacksquare)$ error intervals containing 68 and 95 \% of the realisations (note that Table \ref{tab_hd127756} lists the $3 \sigma$, or 99\%, error interval) centred on the fitted parameter. Below those symbols in each panel are the $1 \sigma~(\blacklozenge)$ errorbars obtained from the formula definition of standard deviation and the mean ($\bullet$) of the distribution with the standard error on the mean. These distributions are shown because the frequencies are the closest to each other.\label{distros_hd127756}}
\end{figure}

A comparison of the closely spaced frequencies near $\sim 1$ c d$^{-1}$ to the window function and to our fit is given in Fig.~\ref{ft_hd127756winft}. Notice that the peak with the largest amplitude has an asymmetric component that is wider than the window function. When that frequency ($\nu_{14}$) is prewhitened, significant power remains near that asymmetry and is fitted as $\nu_{13}$ (shown as the data point with the smallest amplitude in Fig.~\ref{ft_hd127756winft}). These frequencies are spaced by $\sim 0.04$ c d$^{-1}$ which is greater than the Rayleigh criterion for our data ($\sim 0.03$ c d$^{-1}$). The points are clearly separated in frequency within their respective 3$\sigma$ errorbars. The peak labelled as A$_{\textrm{x}}$ ($\nu_{15}$ in Table \ref{tab_hd127756}) is spaced from $\nu_{14}$ at nearly the resolution limit suggested by \cite{loum78} ($\sim 0.06$ c d$^{-1}$).  This peak is clearly resolved from $\nu_{14}$ and has an amplitude that is $\sim 4$ times that of the first side lobe of the window function (labelled as A$_{\textrm{y}}$). Although the amplitude of this peak may be influenced by windowing of the data, we believe the frequencies are resolved.  Frequencies $\nu_{18}$ and $\nu_{19}$ (shown circled in Fig.~\ref{ft_hd127756winft}) have the smallest frequency separation and are barely resolved within their 3$\sigma$ errorbars with a separation of $\sim 0.004$ c d$^{-1}$. The bootstrap distributions for these frequencies are plotted in Fig \ref{distros_hd127756} and shows the parameters are normally distributed and the frequency distributions of $\nu_{18}$ and $\nu_{19}$ nearly overlap. 

We suggest, based on our bootstrap distributions, that frequencies spaced by less than $0.5~T^{-1} \sim 0.0163$ c d$^{-1}$ (within their 3$\sigma$ errorbars) are at the resolution limit of our data set. Using this resolution criterion, frequency pairs   
$\nu_{6}$ and $\nu_{7}$; $\nu_{18}$ and $\nu_{19}$; and $\nu_{23}$ and $\nu_{24}$ should be modelled with caution. We show in section \ref{HD127756mods} that the determination of the rotation frequency of HD 127756 and the general interpretation of the observed variability depends on the frequency ranges and groupings and does not rely on the individual frequencies being fully resolved. In section \ref{DETAILEDmods}, we discuss the limits of detailed modelling of the stars presented in this work.

\subsection{HD 217543}

\begin{figure}[htbp]
\epsscale{.70}
\plotone{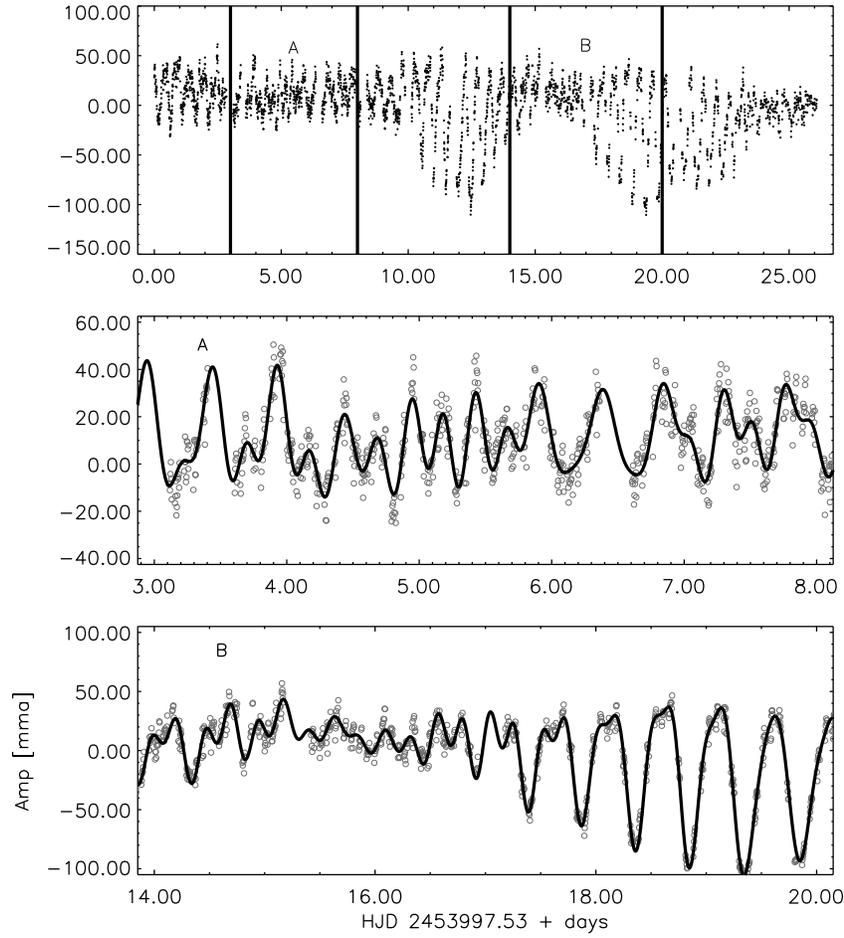}
\caption{{\it MOST} light curve of HD 217543. (Top panel) The full light curve for a total of 26.1 days.The middle and the bottom panels show expanded light curves for the portions A and B (respectively) indicated in the top panel. Solid lines indicate the fit of the 40 most significant frequencies from the frequency analysis of the full light curve (see Table \ref{tab_hd217543}).
\label{lc_hd217543}}
\end{figure}

\begin{figure}[htbp]
\epsscale{.75}
\plotone{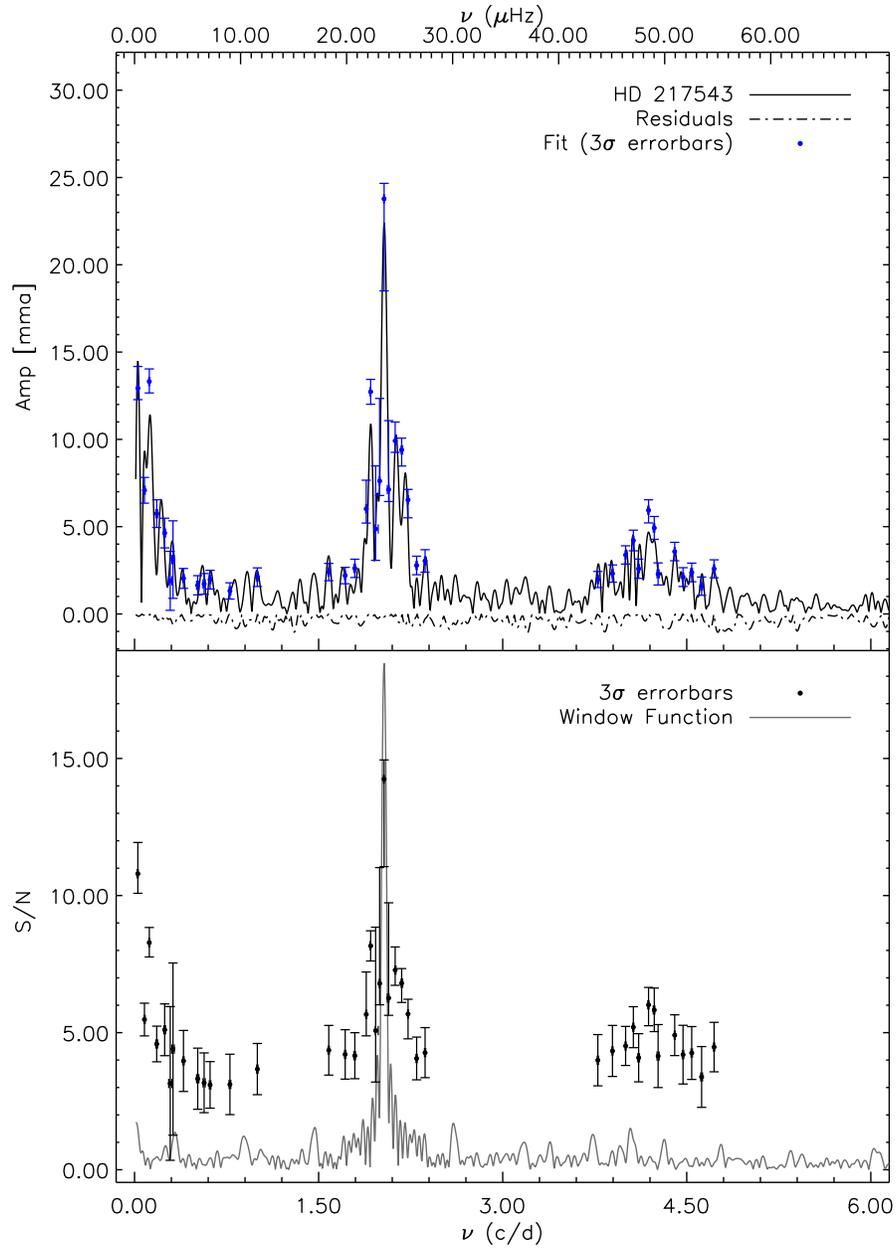}
\caption{Fourier amplitude spectrum of the light curve of HD 217543 and the identified frequency parameters from Table \ref{tab_hd217543}. The panels and the meaning of the symbols are described in Fig.~\ref{ft_hd127756}. \label{ft_hd217543}}
\end{figure}

{\it MOST} observed HD 217543 as a guide star for a total of 26.1 days. Figure~\ref{lc_hd217543} shows the light curve with clear periods of $\sim 0.5$ and $\sim 0.25$ days with modulations characteristic of more complex multi-periodicity. The fit to the 40 most significant frequencies (see Table \ref{tab_hd217543}) is shown in zoomed regions labelled A and B in the lower panels of the plot. Note that in Table \ref{tab_hd217543} there are 6 frequencies with S/N ranging from 3.09 to 3.38. These are below the S/N $\sim 3.5$ limit described above and represent $\gtrsim$ 2$\sigma$ detections \citep{kus97}. They are included to illustrate that within the S/N errors plotted in the lower panel of Figure~\ref{ft_hd217543} all identified frequencies reach the S/N $\sim 3.5$ limit. These periodicities do not adversely influence the fit and do not change the physical interpretation of the observed variations described in the following sections.

Figure~\ref{ft_hd217543} shows an amplitude spectrum of HD 217543 in the top panel and the S/N of the identified frequencies and the window function of the data in the lower panel. As with HD 127756, most frequencies are grouped around three ranges; $\sim 0$ c d$^{-1}$, $\sim 2$ c d$^{-1}$, $\sim 4$ c d$^{-1}$. The second and the third frequency range is higher by a factor of $\sim2$ than the corresponding ones of HD 127756, indicating the rotation frequency of HD 217543 is roughly twice that of HD 127756 (see section \ref{HD217543mods}). Frequencies $\nu{_1}=0.0269$ c~d$^{-1}$ and $\nu{_2}=0.0806$ c~d$^{-1}$ are close to the length of run  ($1/26.1 = 0.0383$ c~d$^{-1}$) but were included to reduce the residuals in the light curve. They may not be intrinsic stellar pulsations.

\begin{figure}[htbp]
\epsscale{1.0}
\plottwo{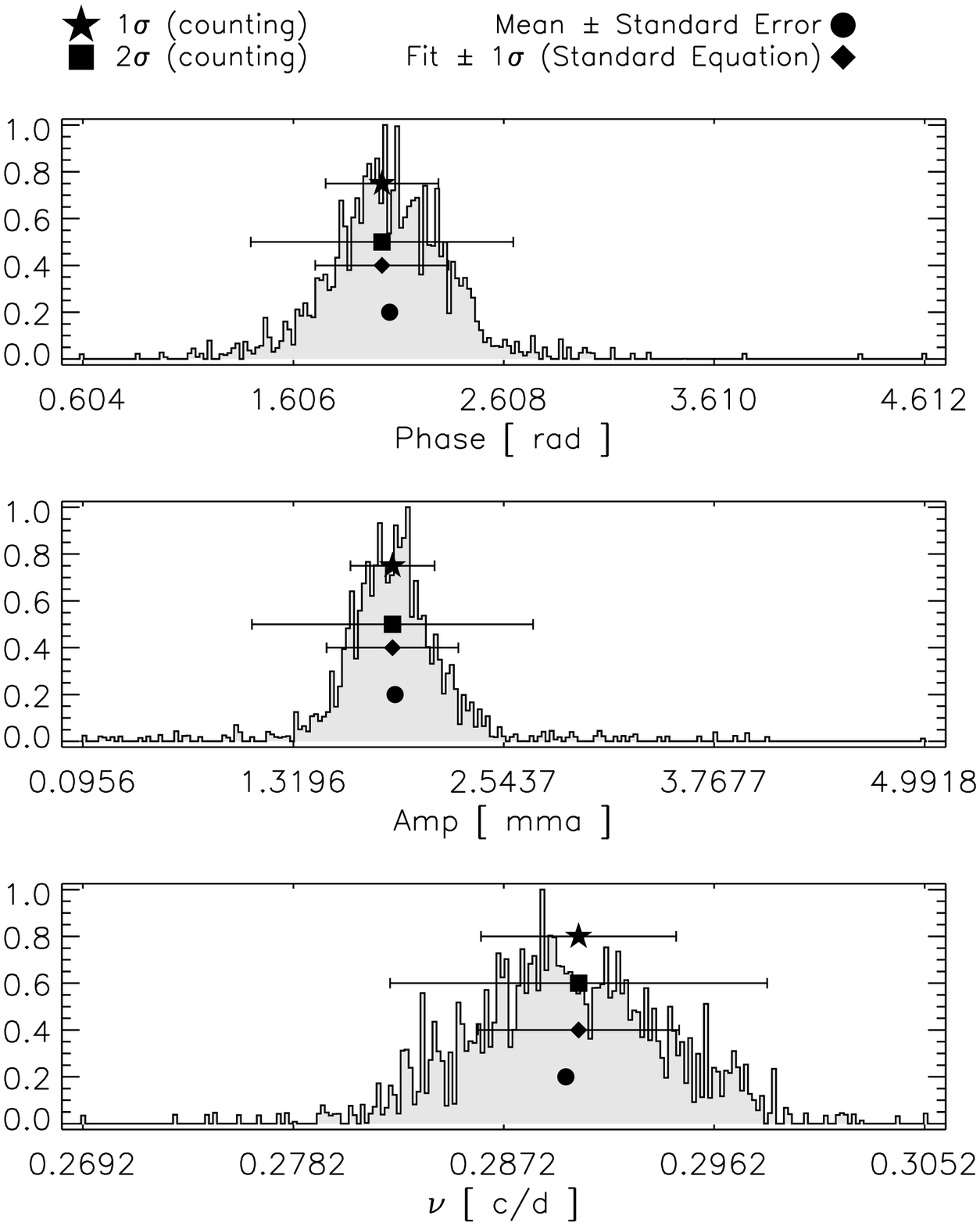}{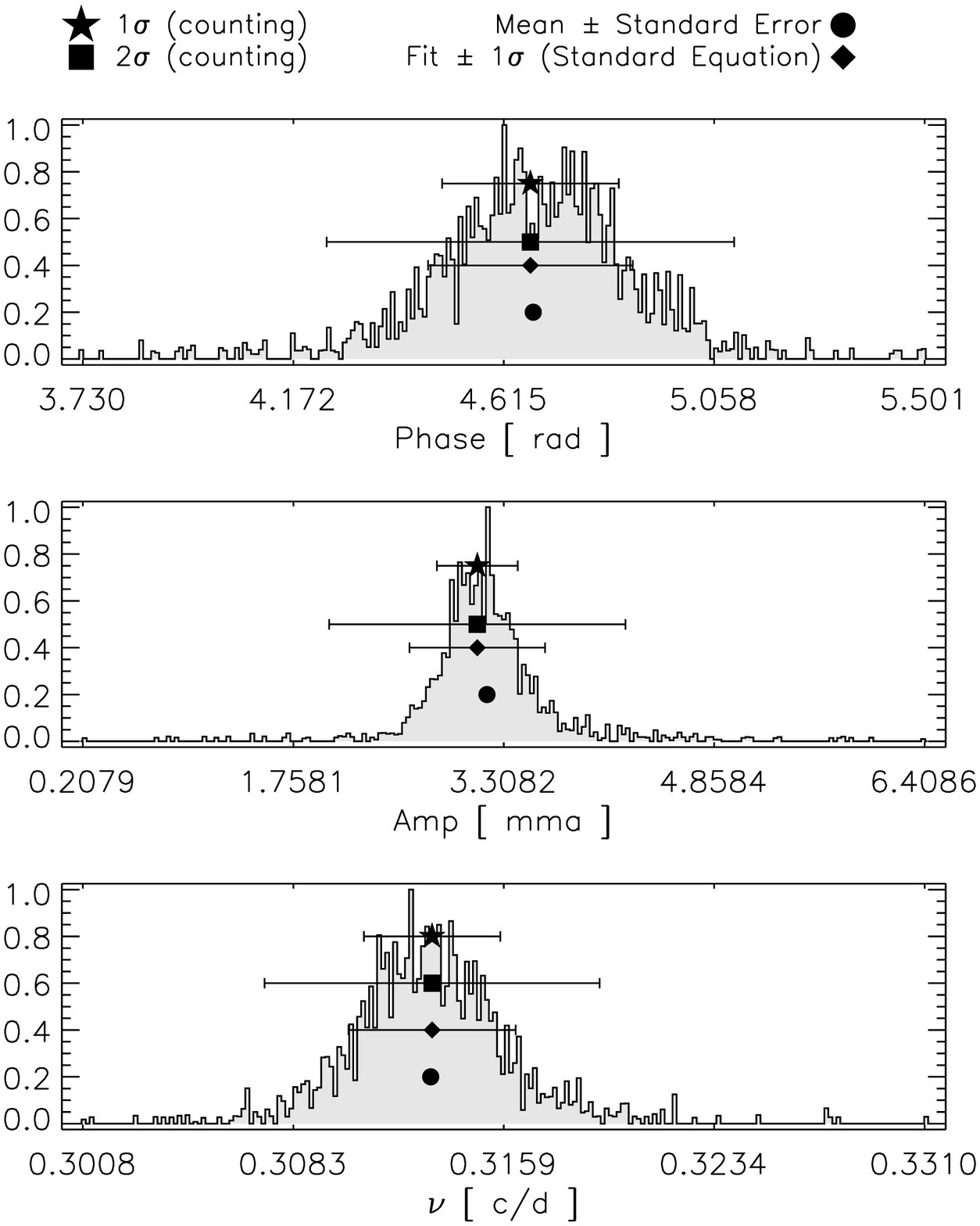}
\caption{A comparison of bootstrap distributions for parameter sets ($\nu_{6}$, $\textrm{A}_{6}$, $\phi_{6}$) and ($\nu_{7}$, $\textrm{A}_{7}$, $\phi_{7}$)  [see Table \ref{tab_hd217543}] for 100000 realisations of the HD 217543 light curve. Symbols are the same as those in Fig.~\ref{distros_hd127756}. These distributions are shown because the fitted frequencies are the closest to each other. In this case, the long tails on the frequency distributions suggest that these frequencies are not fully resolved. \label{distros_hd217543}}
\end{figure}

Frequencies $\nu_{6}$ and $\nu_{7}$ of Table \ref{tab_hd217543} overlap within their 3$\sigma$ uncertainties. The bootstrap distributions are given in Fig.~\ref{distros_hd217543} and show all parameters are normally distributed like those in Fig.~\ref{distros_hd127756} for HD 127756. However, the long tails on the frequency distributions suggest that these frequencies are not fully resolved. If we adopt the same resolution criterion as for HD 127756,  frequencies spaced
less than $0.5~T^{-1} \sim 0.0192$ c d$^{-1}$ (within their 3$\sigma$ errorbars) are at (or below) the resolution limit of our data set. This means frequency sets $\nu_{6}$ and $\nu_{7}$; $\nu_{19}$ and $\nu_{20}$; and $\nu_{34}$ and $\nu_{35}$ should be modelled with care. Once again, the resolution of individual frequencies is not a requirement for the determination of the rotation frequency of this star (see section \ref{HD217543mods}). 

\section{Theoretical models}\label{theory}

The groupings of frequencies seen in the two stars is consistent with high radial-order g-mode oscillations of which frequencies in the co-rotating frame are smaller than the rotation frequency meaning that HD 127756 and HD 217543 are two new SPBe stars. The mean frequencies of the second and third frequency groups for  HD 217543 are about twice the corresponding ones for HD 127756, indicating that the rotation frequency of HD 217543 is about double that of HD 127756 (see below).

The modelling here is the same as that in \citet{wal05b} and \citet{sai07}. The same chemical composition $(X,Z) = (0.7,0.02)$ is assumed for all models. We have considered models computed with OP \citep[Opacity Project;][]{OP} opacity tables as well as models with OPAL(95) opacity  tables \citep{OPAL}, taking into account the recent theoretical results \citep{jef06,jef07,mig07} that OP opacities tend to excite pulsations in hotter models than those with OPAL opacities. The equation of state in the envelope was obtained by solving Saha's equation for Hydrogen, Helium, and Carbon. The structure of a convection zone in the envelope was calculated with a local mixing length theory using a mixing length of 1.5 times the pressure scale height. The perturbation of convective flux was neglected in the stability analysis (described below) and no overshooting from the convective core was assumed.

The stability of nonradial pulsations in rapidly-rotating stars was examined using the method of \citet{lee95}, in which the deformation [proportional to $P_2(\cos\theta)$] of the equilibrium structure due to the centrifugal force is included. The angular dependencies of pulsational perturbations are expanded into terms proportional to spherical harmonics $Y_{l_j}^m$ for a given azimuthal order $m$ ($Y_{l'_j}^m$ for toroidal velocity field) with  $l_j = |m| + 2j$ ($l'_j = l_j+1$) for even modes and $l_j = |m| + 2j+1$ ($l'_j=l_j-1$) for odd modes with $j= 0, 1, \ldots N$. The series is truncated at $N=9$ so we can obtain accurate eigenfunctions for low-degree modes within a reasonable computing time. We adopt the convention that a {\it negative} $m$ represents a {\it prograde} mode (in the co-rotating frame) with respect to the stellar rotation. Even (odd) modes are symmetric (anti-symmetric) with respect to the equatorial plane.  We designate the angular-dependence type of  a mode by a set of ($m$,$\ell$) in which $\ell$ is defined as the $l_j$ value of the largest-amplitude component. Taking into account that high surface degrees reduce the visibility of the modes, we consider in this paper (as in our previous analyses) modes with $\ell \le 2$. 

\subsection{Models for HD 127756}\label{HD127756mods}

\begin{figure}[htbp]
\epsscale{.60}
\plotone{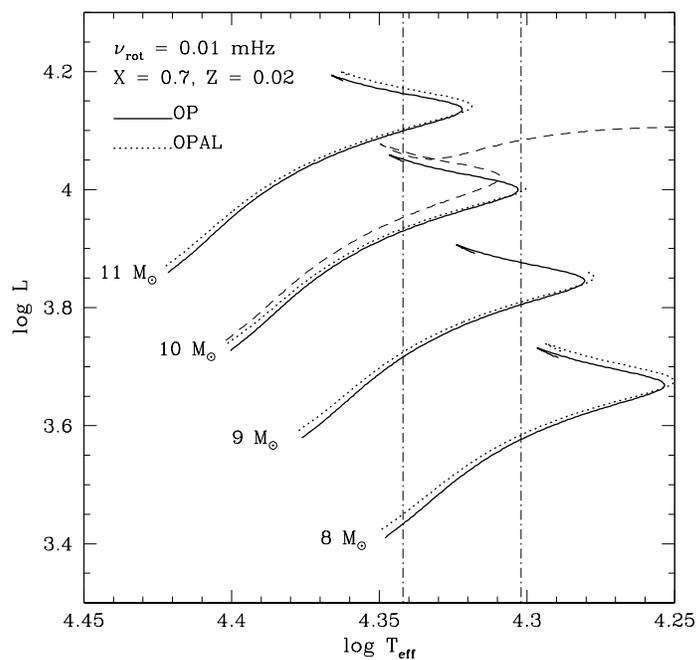}
\caption{
Evolutionary tracks of models computed for HD 127756.  A constant and rigid rotation at a rate of 0.01 mHz (0.86 c d$^{-1}$) is assumed throughout the evolution. The evolutionary track of $10M_\odot$ non-rotating models computed using OPAL opacities is shown with a dashed line. Vertical dash-dotted lines indicate the estimated range in the effective temperature for HD 127756 based on photometric indices. 
\label{tracks_hd127756}}
\end{figure}


\citet{koz85} gives values of $V_0=6.56$, $(B-V)_0 = -0.22$, and $(U-B)_0=-1.01$ for HD 127756. The $(B-V)_0$ value corresponds to $\log T_{\rm eff} = 4.322$ according to Code et al's (1976) calibration. (Note that using \citet{flo77}'s table gives $\log T_{\rm eff} = 4.317$.) Assuming an error of $\pm0.01$ in $(B-V)_0$, which corresponds to $\pm 0.02$ in $\log T_{\rm eff}$, we estimate the effective temperature of HD 127756 lies in the range of $\log T_{\rm eff} = 4.32 \pm 0.02.$ This range is shown by vertical lines on the HR diagram in Fig.~\ref{tracks_hd127756} along with some evolutionary tracks. From a relation between $(U-B)_0$ and $M_{\rm v}$ for Be stars, \citet{koz85} estimated $M_{\rm v}\approx -3.4 \pm 0.4$ mag for HD 127756. Applying a bolometric correction of $-2.1$ mag \citep{cod76,flo77}, we derive a luminosity of $\log L/L_\odot \approx 4.1\pm 0.2$.

An evolutionary series of models was computed for masses of $8,~9,~10$, and $11M_\odot$ because they cover the range of effective temperatures derived above, during the main-sequence evolution phase (Fig.\ref{tracks_hd127756}). Only the 10 and 11$M_\odot$ models cross into the estimated luminosity range during the late stages of main-sequence evolution.  A rotation frequency of $0.01$ mHz (0.86 c d$^{-1}$) is adopted to approximately fit the observed two groups of frequencies with $m=-1$ and $m=-2$ prograde g-modes. 

\begin{figure}[htbp]
\epsscale{.90}
\plotone{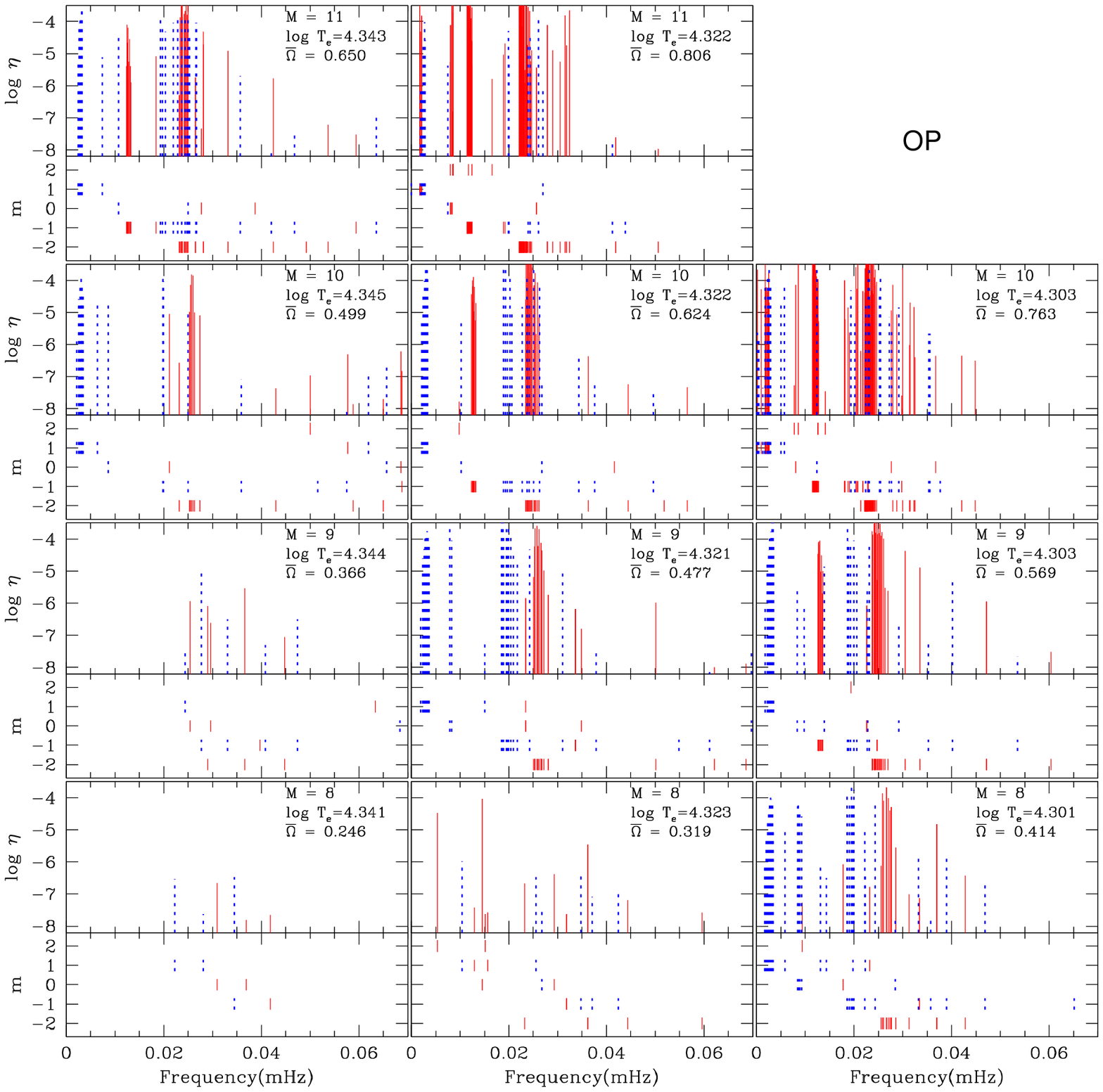}
\caption{
Growth rates $\eta$ and azimuthal order $m$ versus frequencies (in the observers' frame) of excited low-degree ($\ell \le 2$) modes are shown for selected models for HD 127756 computed with OP opacities. Solid (red) lines are for even (symmetric with respect to the equator) modes, while broken (blue) lines for odd modes.
\label{OP_01mhz}}
\end{figure}

\begin{figure}[htbp]
\epsscale{.90}
\plotone{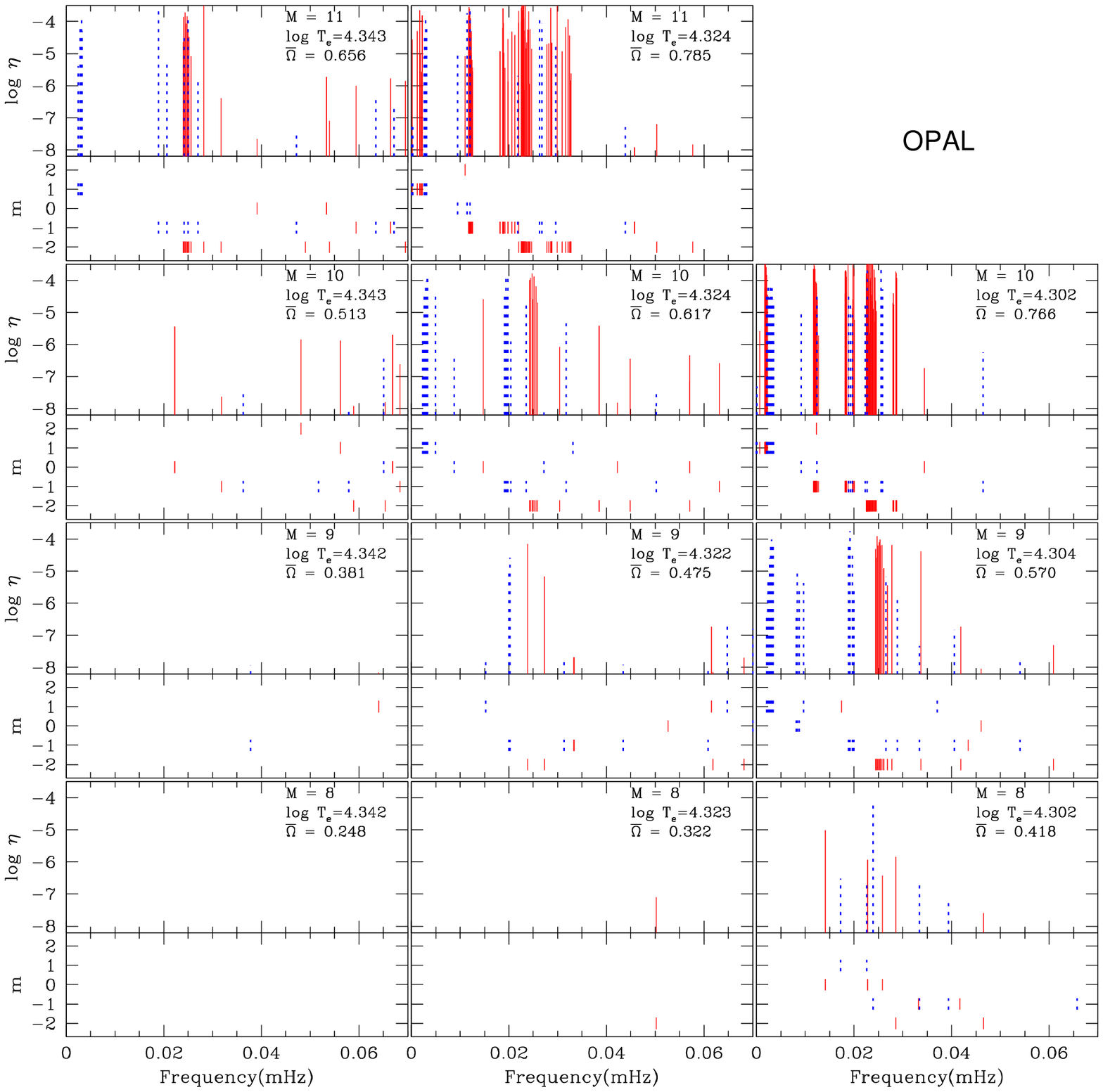}
\caption{
The same as Fig. \ref{OP_01mhz} but for models computed with OPAL opacities.
\label{OPAL_01mhz}}
\end{figure}

The stability analysis of the pulsations has been examined for models having $\log T_{\rm eff} \approx 4.34,~4.32$, and $ 4.30$ for each mass and for both OP and OPAL opacities. The growth rates and $m$ values of excited low-degree ($\ell \le 2$)
modes are shown in Fig.\ref{OP_01mhz} for the OP opacity and in Fig.\ref{OPAL_01mhz} for the OPAL opacity models. Red (full) lines are for modes symmetric with respect to the equatorial plane and blue (broken) lines are for anti-symmetric modes.

Generally, more modes are excited in cooler and more luminous models. By comparing the models with OP and OPAL opacities (Figs~\ref{OP_01mhz},~\ref{OPAL_01mhz}), a similar number of modes are excited in cooler and more luminous models using the OPAL opacities compared to the cases that use OP opacities. This suggests that the stability boundary seems to shift to redder and more luminous values for the OPAL opacities, which is consistent with the result of the stability analysis by \citet{mig07} for non-rotating B-stars.   

Since both high-order g-modes (and r-modes) and some low-order g-modes are excited in these models, the 
frequency versus growth-rate diagrams are more complex than those for less massive models of HD 217543 and HD 163868 (see below). We note that some of the excited low-order g-modes have considerable contributions from high $l$ components which tend to reduce their visibility in integrated light.

In order to be consistent with the observed frequencies of HD 127756 (Fig.~\ref{ft_hd217543}), at least two groups
of frequencies around 0.011 mHz (1 c d$^{-1}$) and 0.023 mHz (2 c d$^{-1}$) should be excited. Among the models with the OP tables shown in Fig.~\ref{OP_01mhz}, this requirement is met by models of mass $11 M_\odot$, the cooler two models of mass $10 M_\odot$, and the coolest model of $9 M_\odot$. On the other hand, among the models with the OPAL opacity tables (Fig.\ref{OPAL_01mhz}), the cooler $11 M_\odot$ model and the coolest $10 M_\odot$ model are more or less consistent with HD 127756. These models (except for the $9M_\odot$ model) are luminous enough to be consistent with the range of the luminosity estimated above. 

Each panel of Figs.~\ref{OP_01mhz}, and \ref{OPAL_01mhz} gives the value of  the normalized rotation frequency, $\overline{\Omega}\equiv \Omega/\sqrt{GM/R^2}$ corresponding to $\Omega = 0.01$ mHz. Since the radius $R$ refers to a mean radius (which is smaller than the equatorial radius), the critical rotation on the equator corresponds to $\overline{\Omega}\approx 0.75$. The values of $\overline{\Omega}$ in these models indicate that the equatorial rotation speeds on the surface are not far from the critical speed, which seems common in Be stars.  Although the angular rotation speed of HD 127756 is much smaller than that of HD 163868 \citep[0.016 mHz][]{wal05b}, the larger radius of  HD 127756 makes the surface rotation velocity near critical. 

\begin{figure}[htbp]
\epsscale{.70}
\plotone{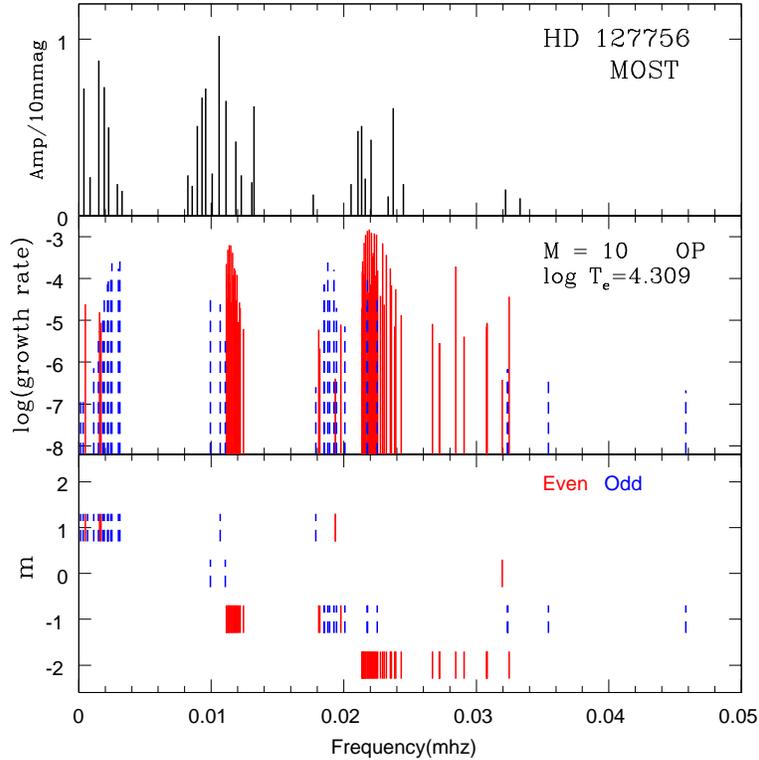}
\caption{
Observed frequencies of HD 127756 are compared with a closely matched $10 M_\odot$ model with OP opacities. The bottom and the middle panels show azimuthal orders and growth rates of excited modes with $\ell \le 2$. The top panel shows observed frequencies and corresponding amplitudes. A slightly smaller rotational frequency of 0.0094 mHz ($\overline\Omega = 0.67$) was assumed when converting frequencies in the co-rotating frame to those in the inertial frame to have a better match. 
\label{besthd127756}}
\end{figure}

Figure \ref{besthd127756} provides a comparison between the observed frequencies with a closely matched model of $10M_\odot$ using OP opacities. Most of the excited frequencies in the lowest frequency group ($<0.004$ mHz) are odd r-modes of $m = 1$. Some of the frequencies in this group are retrograde, even g-modes of $m=1$, in which high $l_j$ components contribute significantly to the eigenfunction. The observed frequency group at 0.011 mHz ($\sim 1$ c d$^{-1}$) is mainly identified with prograde, high-order (n = $23 - 41$) g-modes of $m=-1$ that are symmetric with respect to the equatorial plane (i.e., even modes). Only a few odd g-modes contribute to the group. The frequency range actually observed for this group is still larger than the predicted range. The frequency group at 0.023 mHz ($\sim 2$ c d$^{-1}$) is mainly covered by prograde even g-modes of $m=-2$ with radial orders ranging from 26 to 55. The even $m=0$ mode at 0.032 mHz ($\approx 2.8$c d$^{-1}$) is the 5th radial order g-mode with a dominant $l_j=2$ component. The $m=-1$ odd mode at a similar frequency is the 7th radial order g-mode with a dominant $l_j=2$ component. The visibility of relatively high frequency ($> 0.026$ mHz, 2.25 c d$^{-1}$) $m=-2$ modes is probably low because contributions to the eigenfunctions from high $l_j$ components are large in these modes.


\subsection{Modelling HD 217543}\label{HD217543mods}

\citet{zor05} estimated the parameters of HD 217543 as $\log T_{\rm eff} = 4.270$, $\log g = 3.95$, and $M=6.8M_\odot$. These values yield $\log L/L_\odot = 3.355$. We have calculated evolutionary models for masses
of 6, 7,  and $8M_\odot$ rotating with a frequency of 0.02 mHz (1.73 c d$^{-1}$). The rotation frequency was chosen so that high-order g-modes have frequencies in the observers' frame consistent with the observed oscillation frequencies for HD 217543 (see below). Figure \ref{hd217543tracks} shows the calculated evolutionary tracks with HD 217543 (a big circle) put beside the $7M_\odot$ tracks on the HR diagram in accordance with \citeauthor{zor05}'s parameters.

\begin{figure}[htbp]
\epsscale{0.6}
\plotone{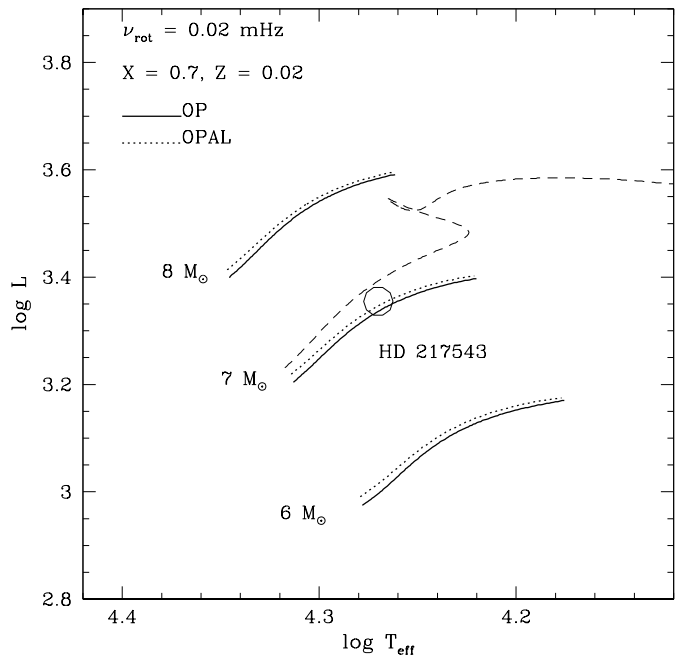}
\caption{
The position of HD 217543 is shown by a circle along with evolutionary tracks for 6, 7, and $8M_\odot$
models rotating at a rate of 0.02 mHz. The dashed line shows the evolutionary track of $7M_\odot$ non-rotating models calculated with OPAL opacities. \label{hd217543tracks}}
\end{figure}

\begin{figure}[htbp]
\epsscale{0.7}
\plotone{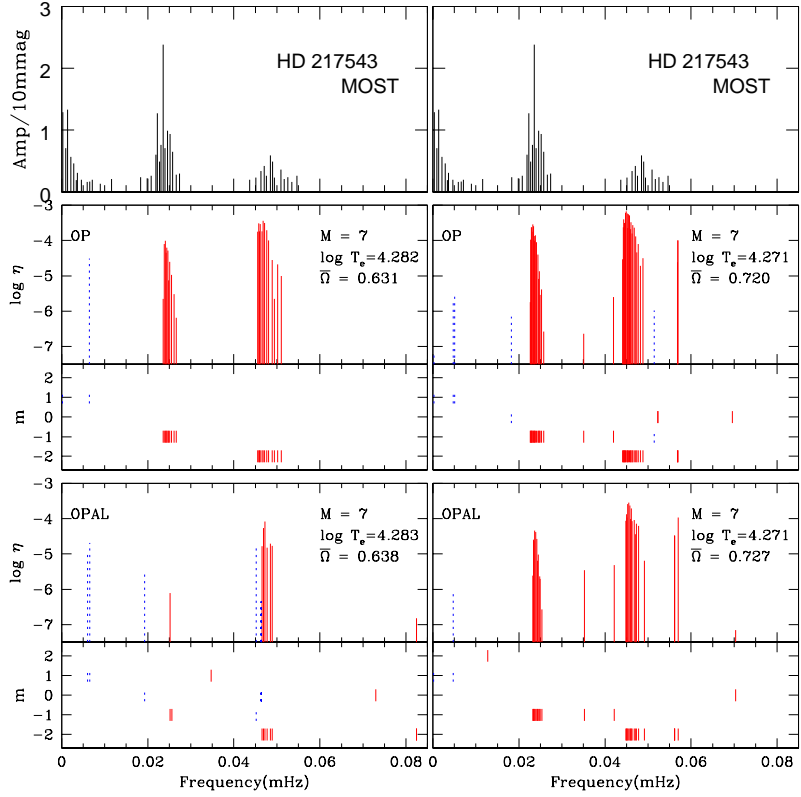}
\caption{Growth rates and azimuthal order $m$ versus pulsation frequencies of excited modes with $\ell \le 2$ for rapidly rotating $7 M_\odot$ models computed with the OP (middle panels) and OPAL (bottom panels) opacities. A rotation frequency of 0.02 mHz (1.728 c d$^{-1}$) is assumed. Red lines are for even modes symmetric with respect to the equatorial plane, while blue (broken) lines for the anti-symmetric modes (odd modes). For comparison, an observed amplitude-frequency diagram of HD 217543 is shown in both of the top panels.}
\label{opopalm7}
\end{figure}

Since the parameters of HD 217543 are relatively well determined, we present the results from a pulsation analysis for models with a mass $7M_\odot$ and appropriate effective temperatures. Figure \ref{opopalm7} shows growth rates and azimuthal order $m$ (prograde modes correspond to $m<0$) versus pulsation frequencies of excited modes in $7M_\odot$ models rotating at a rate of 0.02 mHz (1.7 c d$^{-1}$). The top and the bottom panels are for models with OP and OPAL opacities, respectively. The cooler models have $\log T_{\rm eff} = 4.271$; close to the value estimated by \citet{zor05}. Results for slightly hotter models ($\log T_{\rm eff} \approx 4.282$) are also shown to exhibit the dependence of mode excitation on the effective temperature. The equatorial rotational velocities are about 390 km~s$^{-1}$ for the cooler models and about 360 km~s$^{-1}$ for the hotter models, indicating the inclination angle of the rotation axis is very high ($70^\circ-90^\circ$). We expect only symmetric (even) modes to be detected when observing the star at such a high inclination angle.

Prograde, high-order g-modes of $m=-1$ and $-2$ have frequencies of about $0.024-0.025$ mHz and about $0.045-0.05$ mHz in the inertial frame. The predicted frequency groups agree well with those detected by \textit{MOST} in HD 217543. As is the case for the HD 127756 models, the OP opacities excite a larger number of g-modes compared with OPAL opacities, and among the models shown, those with OP opacities agree better with observed frequencies. However, these models cannot explain the very low frequencies ($< 0.01$ mHz) we observed since very few r-modes are excited and these odd modes should be invisible in the nearly equator-on orientation.

\section{Discussion}
HD 127756 and HD 217543 join HD 163868 \citep{wal05b}, and $\beta$ CMi \citep{sai07} as members of the SPBe class.  We can start to investigate this class in more details with four stars.

\subsection{Revisiting HD 163868}\label{hd163868mods}

\citet{wal05b} considered HD 163868 as a B5 Ve star \citep{1973MmRAS..77..199T}, and compared the observed frequencies with a $6 M_\odot$ model having $\log T_{\rm eff}\approx 4.23$ and rotating at a frequency of $0.016$ mHz ($\Omega/\sqrt{GM/R^3}=0.60$). However, there are some observational facts that indicate HD 163868 is hotter than B5 V. The Michigan Catalog of HD stars \citep{hou82} assigns B2/B3 V:ne to the star, while \citet{gar77} have classified it as B1.5 V:ne. \citet{koz85} obtained $(B-V)_0 = -0.19$ and $(U-B)_0 = -0.82$ with $E(B-V) = 0.19$. The $(B-V)_0$ value corresponds to B3.5 and $\log T_{\rm eff} = 4.251$ \citep{flo77}, while the $(U-B)_0$ value corresponds to B2 and $\log T_{\rm eff} = 4.331$ \citep{lan92}. Furthermore, \citet{ste98} obtained $b-y=0.056$, $c_1=0.140$, and the Geneva emission-free $\beta$ index $\beta_c = 2.630$ (mean value) for HD 163868. Adopting the relations $E(b-y)=0.74E(B-V)$ \citep{dav77} and $c_0 = c_1 - 0.24E(b-y)$ \citep{cra75} yields $c_0 = 0.106$. Substituting these values for $\beta_c$ and $c_0$ into the interpolation formula for $\log T_{\rm eff}$ derived by \cite{bal84} we obtain $\log T_{\rm eff}= 4.314$. This corresponds to a main-sequence spectral type of around B2 \citep{gra94}.

\begin{figure}[htbp]
\epsscale{0.7}
\plotone{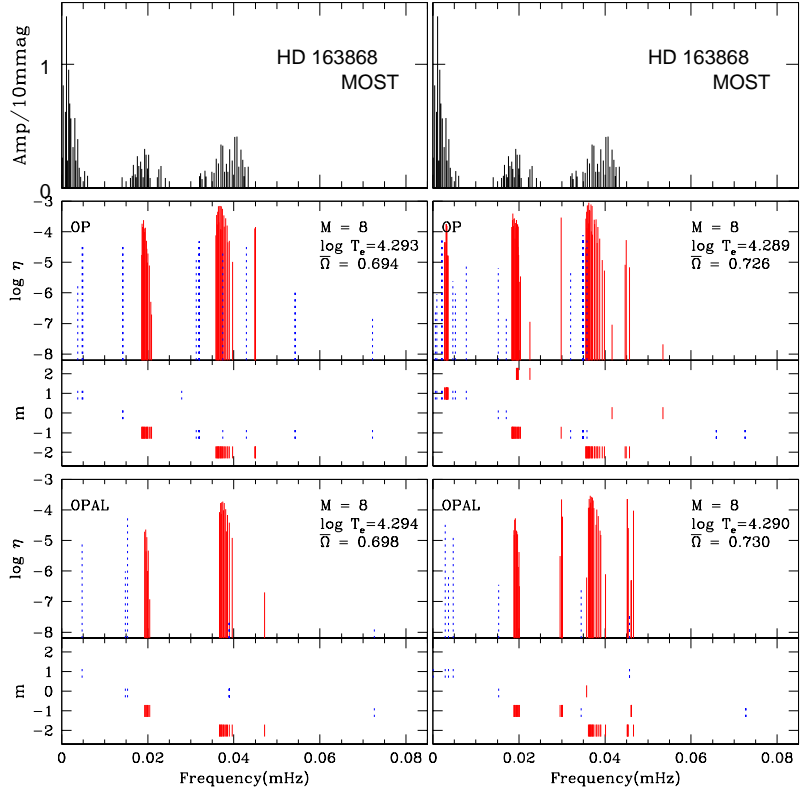}
\caption{
The growth rates and azimuthal orders versus frequencies of excited nonradial pulsations ($\ell \le 2$) for $8 M_\odot$ main-sequence models rotating at a rate of $0.016$ mHz. The top panels show observed amplitude versus frequency of HD 163868 as observed by {\it MOST}.  
\label{hd163868}}
\end{figure}

The parameters referenced above indicate that the effective temperature of HD 163868 should be $\log T_{\rm eff}\approx 4.30$ which corresponds to B2 V rather than B5 V. Therefore, we have re-modelled HD 163868 by adopting
a mass of $8 M_\odot$. We have assumed a rotation frequency of $0.016$ mHz as before. Figure~\ref{hd163868} shows the growth rates and azimuthal orders versus frequencies of excited nonradial pulsations for two models having slightly different $\log T_{\rm eff}$ calculated with OP opacities and two models with OPAL opacities.
The cooler model with OP opacities reproduces well three observed frequency groups. The agreement of this model to the observed data is comparable with that of the old $6 M_\odot$ model shown in \citet{wal05b}. 
The new model, however, rotates nearly critically ($\Omega/\sqrt{GM/R^3}=0.73$).  

Figure~\ref{hd163868} indicates that in the cooler model with OP opacities some retrograde, high-order g-modes of $m=1$ are excited in addition to odd r-modes in the very low frequency range. For these retrograde g-modes, the eigenfunctions are significantly affected by high $l_j$ components, which indicates that the visibility of these modes should be low (This property has already been pointed out for the model of HD 127756 shown in Fig.~\ref{besthd127756}). \citet{dzi07} and \citet{sav07} argued that these retrograde g-modes of $m=1$ are responsible for the very low frequencies observed in HD 163868.
In contrast to those results, only a few such modes are excited in our analysis \citep[no such modes were found excited in our previous models in][]{wal05b}. 

The difference might be explained by the fact that in our method including the effects of centrifugal deformation yields 
stronger damping for retrograde g-modes (Lee 2008; private communication), and that our analysis is restricted to low surface-degree modes of $\ell \le 2$ while high $l_j$ components tend to be significant for retrograde g-modes. Further theoretical and observational investigations are needed to clarify the nature of  the very low frequencies.

\subsection{Rotation frequencies of SPBe stars}

\begin{figure}[htbp]
\epsscale{0.7}
\plotone{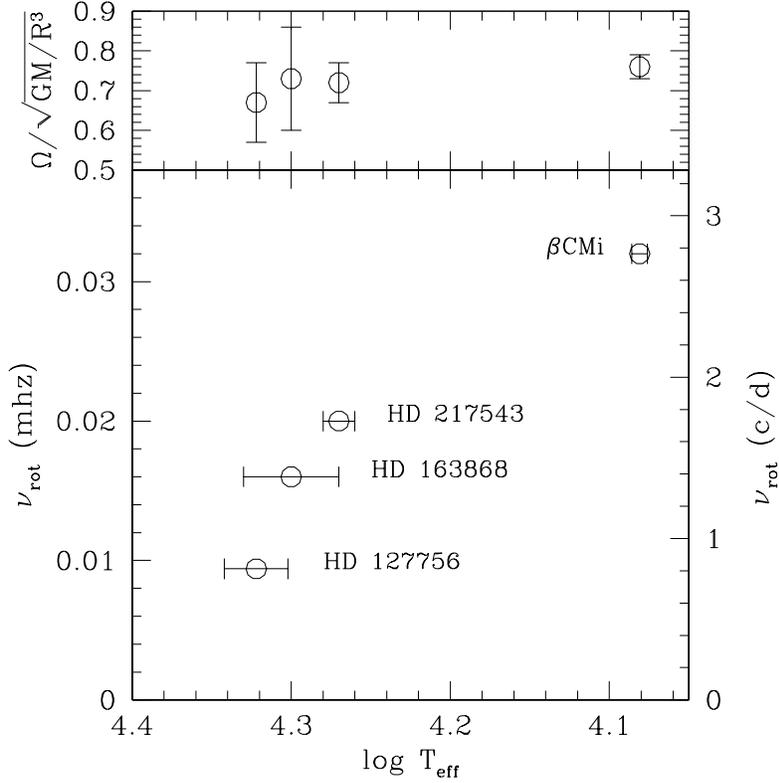}
\caption{The estimated rotation frequencies versus effective temperatures for the four SPBe stars observed by the {\it MOST} satellite (bottom panel). The top panel shows normalized rotation frequencies $\overline{\Omega}\equiv\Omega/\sqrt{GM/R^3}$, where $R$ is the mean radius. The critical rotation at the equator occurs when $\overline{\Omega}\approx 0.75$. Probable errors in $\overline{\Omega}$, which come from uncertainties in the stellar radius, are estimated as follows: for $\beta$ CMi and HD 217543 $\Delta \log R \approx |2\Delta \log T_{\rm eff}|$; for HD 163868 the difference from the previous model \citep{wal05b} is adopted as a probable error; and for HD 127756 an error of $\pm0.1$ is estimated from models shown in Figs.~\ref{OP_01mhz} and \ref{OPAL_01mhz}. \label{te_rot}}
\end{figure}

As we demonstrated in section \ref{theory}, comparing the observed oscillation frequencies of a SPBe star with theoretical models yields the rotation frequency of the star without referring to $v\sin i$. Using our rotational frequencies, we can derive the equatorial velocity of each star and see how close it is to the critical velocity if an accurate estimate for the equatorial radius is available.

The {\it MOST} satellite has detected the SPBe-type variations in four stars so far: HD 127756 and HD 217543 (this paper), 
HD 163868 \citep{wal05b}, and $\beta$ CMi \citep{sai07}. Figure~\ref{te_rot} shows the rotation frequencies derived for these stars as a function of effective temperature (bottom panel). This figure shows that the rotation frequency decreases systematically as the effective temperature increases. This is due to the fact that the hotter (more massive) Be stars have
larger radii. The top panel of Fig.~\ref{te_rot} shows the rotation frequency normalized as $\overline{\Omega}=\Omega/\sqrt{GM/R^3}$, where $R$ is the mean radius taken from the best model. The normalized rotation frequencies lie between 0.7 and 0.8 for the four cases indicating that these stars rotate nearly critically at the equator. Since the equatorial radius is larger than the mean radius $R$, the critical value of $\overline{\Omega}$ is $\approx 0.75$ according to the Roche model. Although the rotation frequency itself is well determined, the ratio to the critical rotation frequency is affected by uncertainty in the stellar radius\footnote{In our analysis the deformation from the centrifugal force was included up to the order $\Omega^{2}$. It is worth noting that this approximation is not accurate for the equatorial region of a nearly critically rotating star, and might affect the stability of g-modes. However, a shift of frequency ranges of excited g-modes in the co-rotating frame would not change our conclusions.}. 

Knowing how rapidly a Be stars rotates is an important property in understanding the mechanism of mass ejection from these stars. From a statistical analysis \citet{cra05} found that late-type Be stars tend to rotate nearly critically, while the ratios of the rotation speed to the critical velocity for early type ($T_{\rm eff} \gtrsim 21,000$K) is roughly uniformly spread from $0.4-0.6$ to unity.  On the other hand, \cite{tow04} concluded that nearly all Be stars could be rotating nearly critically.   

As discussed in \citet{sai07}, the late-type Be star $\beta$ CMi rotates nearly critically. The values of $\overline{\Omega}$ for early type Be stars in Fig.~\ref{te_rot} look slightly smaller than the value for $\beta$ CMi, but not significantly as low as claimed by \citet{cra05}. We need more observations of SPBe-type oscillations for other Be stars as well as accurate stellar parameters in order to better understand the connection between rotation speed and the Be phenomena.

\subsection{Comparison of g-mode frequencies}\label{DETAILEDmods}

In the previous sections we compared theoretical models with observations mainly with respect to the observed frequency ranges of excited modes rather than the frequencies of individual modes. If the models become good enough, it will be possible to compare each frequency or frequency spectrum of g-modes to observed periodicities to obtain useful information on stellar structure. As a first step for SPBe stars, we present in this subsection exploratory comparisons of g-mode frequencies in the corotating frame. 

\begin{figure}[htbp]
\epsscale{0.6}
\plotone{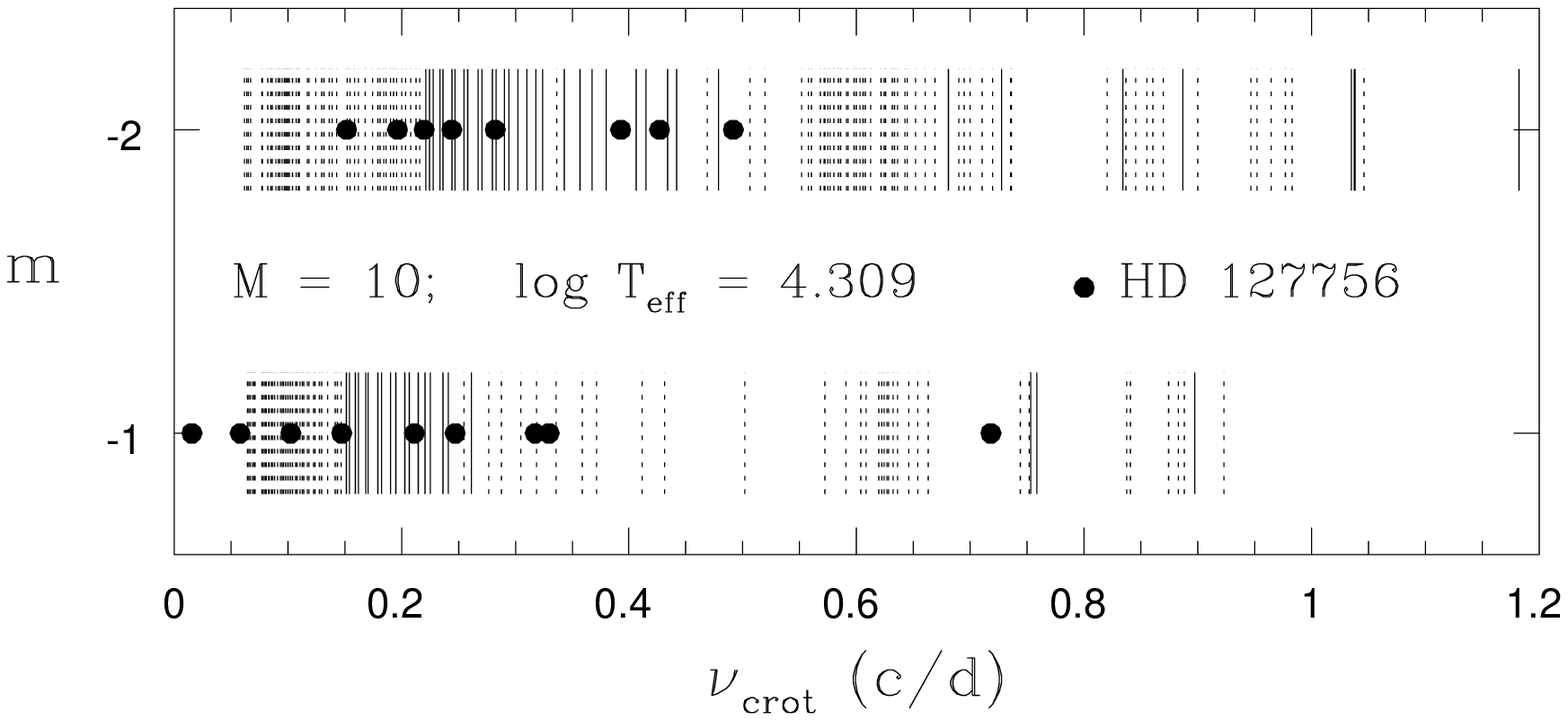}
\plotone{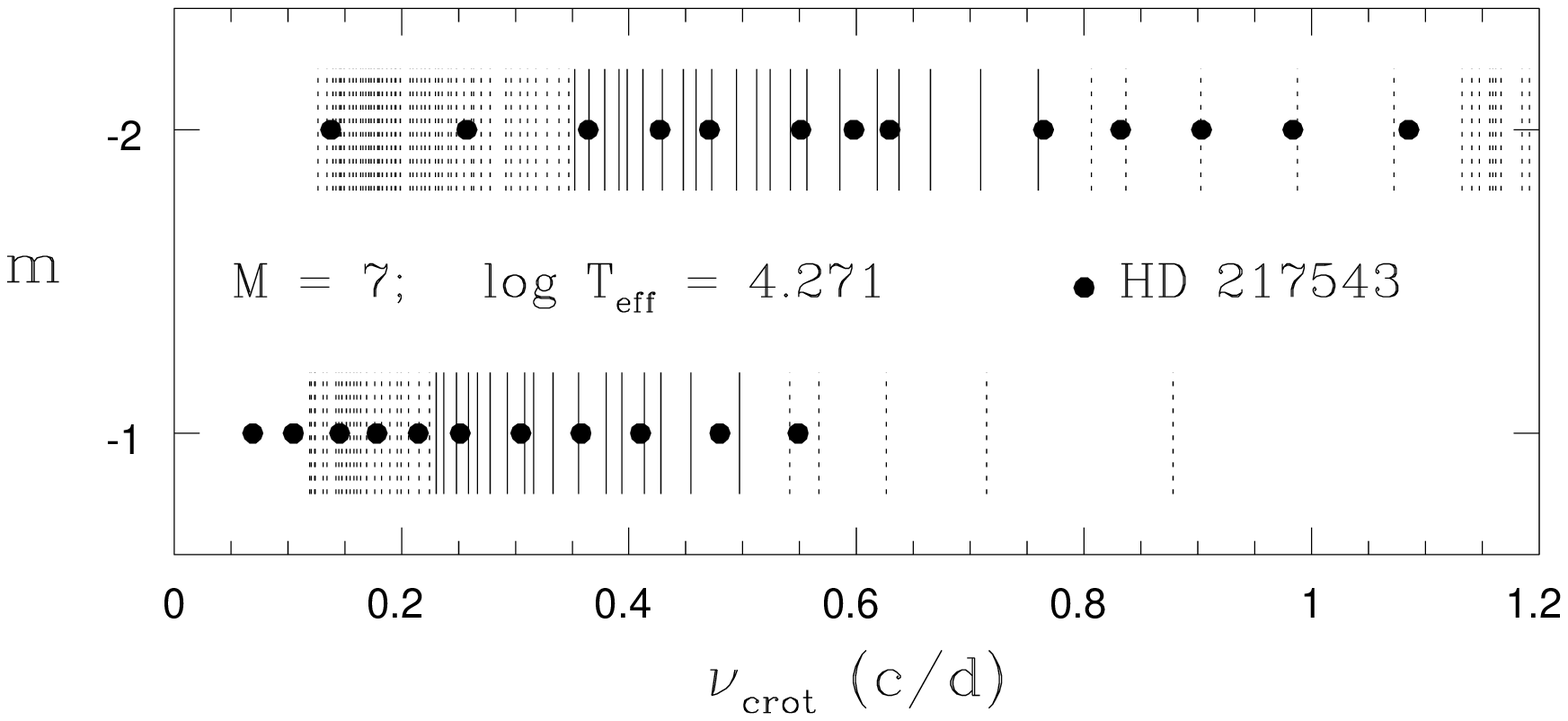}
\caption{Exploratory comparisons of g-mode frequencies in the co-rotating frame. The top panel shows frequencies for HD 127756 compared to the $10 M_\odot$ model shown in Fig.~\ref{besthd127756}. The lower panel plots frequencies for HD 217543 compared to the cooler $7 M_\odot$ model with the OP opacity shown in Fig.~\ref{opopalm7}. Observed frequencies are converted to co-rotating frame frequencies assuming that the second and the third frequency groups of each star belong to prograde modes with $m=-1$ and $m=-2$, respectively. The rotation frequency assumed is $0.0094$ mHz (0.81 c~d$^{-1}$) for HD 127756, while a slightly larger value of 0.021 mHz is adopted for HD 217543
(1.814 c d$^{-1}$) to improve the match. The horizontal axis is the frequency in the co-rotating frame and
the vertical axis indicates azimuthal order $m$. The frequencies of excited and damped g-modes are indicated by solid and dotted bars, respectively.}\label{freq_comp}
\end{figure}

The top panel of Fig.~\ref{freq_comp} compares frequencies of HD 127756 in the co-rotating frame with prograde g-modes in the $10 M_\odot$ model shown in Fig.~\ref{besthd127756}, where the second and third observed frequency
groups are assumed to be prograde g-modes with $m=-1$ and $m=-2$, respectively. Theoretical frequencies of excited (damped) modes are indicated by solid (dotted) vertical lines, while observed frequencies are indicated by large dots.
In this model, excited g-mode groups have radial orders of $n=23 - 41$ for $m=-1 ~(\ell = 1)$ and $n=26 - 55$ for $m=-2 ~(\ell = 2)$. Generally, frequency spacings of excited g-modes are smaller than observed frequency spacings. The former tend to be even smaller than the observational limit for this data set, which is of order $\sim 1/(30.7{\rm days})$. The dense spacings of damped modes around $\sim 0.6$ and $\sim 0.9$ c~d$^{-1}$ breaks the usual g-mode frequency spectrum (where the spacing should increase with frequency) because of significant contributions from high $l_j$ components.

The bottom panel of Fig.~\ref{freq_comp} compares HD 217543 to the $7 M_\odot$ model with OP opacity shown in
the right panel of Fig.~\ref{opopalm7}. Observed frequencies are converted to the co-rotating frame by assuming the rotation frequency is 0.021 mHz (to have a better agreement we have employed a slightly higher rotation frequency than before). Excited g-modes in this model have radial orders of $n = 13 - 29$ for $m=-1 ~(\ell = 1)$, and  $n=17 - 37$ for $m=-2 ~(\ell = 2)$. The radial orders tend to be lower and the frequency spacings larger than in the HD 127756 model. In the frequency range of $0.75$ c~d$^{-1}$ $\lesssim \nu_{\rm crot} \lesssim 1.1$ c~d$^{-1}$ for HD 217543, five observed frequencies agree well with model frequencies of $m=-2$ g-modes although most of them are damped modes. In other frequency ranges, however, theoretical frequency spacings for most of the exited modes are smaller than the observed spacings which are limited by the length of the observations ($\sim 1/26.1 \approx 0.038$ c d$^{-1}$).

It is obvious from Fig.~\ref{freq_comp} that agreement between the models and observations is unsatisfactory. To resolve g-mode frequency spacings, observations with much longer baseline would be necessary. In addition, models including possible differential rotations might be necessary to fit with observed frequencies. We are confident that in the near future detailed g-mode asteroseismology should be a possibility for Be stars.

\section{Conclusion}

Precise photometry by the \textit{MOST} satellite has revealed high-order g-mode pulsations in two more rapidly rotating Be stars; HD 127756 and HD 217543. High radial order g-modes with pulsation frequencies in the co-rotating frame that are much smaller than the rotation frequency appear in groups depending on the azimuthal order $m$ in an observational amplitude-frequency diagram. Theoretical models indicate that, in rapidly rotating stars, high-order g-modes are excited near the Fe opacity bump at $T\sim2\times10^5$K as in SPB stars. One difference from slowly rotating SPB stars is the fact that among the high-order g-modes, prograde modes are predominantly excited. These modes have frequencies of $\sim|m|\Omega$ in the observers' frame with $\Omega$ being the rotation frequency. For $m= -1$ and $-2$, expected frequencies are $\Omega$ and $2\Omega$ consistent with observed frequencies of HD 127756 and HD 217543 as well as those of previously discovered in HD 163868 \citep{wal05b}.

An SPBe star provides an opportunity to determine the rotation frequency without referring to $v\sin i$.
We have determined rotation frequencies of $\approx 0.01$ mHz ( $\sim$0.9 c d$^{-1}$) and $\approx 0.02$ mHz ($\sim$2 c d$^{-1}$) for HD 127756 and HD 217543, respectively. Combining these results with previously determined rotation frequencies for HD 163868 and $\beta$ CMi, we have found that the rotation frequencies of Be stars systematically decrease with increasing effective temperature (or increasing the stellar radius). This indicates that the rotation velocity of Be stars stays close to the critical value independently of the effective temperature.

Further observations of SPBe stars are needed to provide details of the properties of rotation velocities among the Be stars. 
In order to do a detailed comparison of g-mode spectra (g-mode asteroseismology) between models and observations it is necessary to observe the SPBe stars at different epochs to both increase the frequency resolution and to confirm the observed periodicities. We expect that such detailed analysis will become possible and provide information about the interior structure of the Be stars in the near future. 

\acknowledgments
This research has made use of the SIMBAD database, operated at CDS, Strasbourg, France. The Natural Sciences and Engineering Research Council of Canada supports the research of C.C., D.B.G., J.M.M., A.F.J.M., J.F.R., and S.M.R.. H.S. is supported by the 21st Century COE programme of MEXT, Japan. R.K. is supported by the Canadian Space Agency. W.W.W. is supported by the Austrian Space Agency and the Austrian Science Fund (P17580-N2). A.F.J.M. is also supported by FQRNT (Qu\'ebec) and C.C is partially supported by a Walter C. Sumner Memorial Fellowship.

\clearpage

\begin{tiny}
\begin{deluxetable}{lcc}
\tablecolumns{3}
\tablewidth{0pt}
\tablecaption{Summary of the {\it MOST} observations of HD 127756 and HD 217543 \label{obssumtab}}
\tablehead{\colhead{} & \colhead{HD 127756} & \colhead{HD 217543}} 
\startdata
Spectral type  &  B1/B2 Vne & B3 Vpe \\
V [mag]  & 7.59 & 6.56 \\
Dates (2006) & May 5 -- Jun 5 & Sep 19 -- Oct 15 \\
Duration [days] & 30.7  & 26.1  \\
Duty cycle [\%] & $\sim 35$ & $\sim 33$ \\
Exposure time [sec] & 0.52  & 1.5  \\
Stacked exposures & 19 & 10 \\
Sampling time [sec]  &  20  & 22 \\
Scatter in unbinned data [mmag] & 2.4 & 7.9 \\
Orbital mean error [mmag] & 0.35 & 3.6 \\ 
\enddata
\end{deluxetable}
\end{tiny}

\clearpage

\begin{tiny}
\begin{deluxetable}{c c c c  c cc cc cc}
\tablecolumns{11}
\tablewidth{0pc}
\tablecaption{HD 127756 periodicities from {\it MOST} photometry \label{tab_hd127756}}
\tablehead{
 \# &$\nu$ [c $\textrm{d}^{-1}$] &A [mmag] &$\phi$ [rad] &$\sim$ S/N &$\sigma_{\nu}$ &
 $3\sigma_{\nu}$ &$\sigma_{A}$ &$3\sigma_{A}$ &$\sigma_{\phi}$ &$3\sigma_{\phi}$
 }
\startdata
  1&   0.0335&      7.2&     3.60&     8.68& $\pm~$   0.0009&   0.0023&      0.2&      0.6&     0.11&     0.30\\
  2&   0.0739&      2.2&     0.59&     6.97& $\pm~$   0.0020&   0.0055&      0.1&      0.4&     0.22&     0.58\\
  3&   0.1300&      8.8&     3.65&     9.50& $\pm~$   0.0007&   0.0017&      0.1&      0.4&     0.07&     0.19\\
  4&   0.1664&      7.3&     1.27&     8.31& $\pm~$   0.0004&   0.0010&      0.2&      0.4&     0.06&     0.16\\
  5&   0.1957&      5.0&     2.64&     8.71& $\pm~$   0.0007&   0.0018&      0.2&      0.5&     0.10&     0.24\\
  6&   0.2530&      1.8&     4.11&     5.12& $\pm~$   0.0031&   0.0078&      0.2&      0.4&     0.32&     0.81\\
  7&   0.2827&      1.4&     4.27&     3.89& $\pm~$   0.0032&   0.0079&      0.2&      0.4&     0.34&     0.88\\
  8&   0.7108&      2.3&     3.31&     5.74& $\pm~$   0.0015&   0.0038&      0.2&      0.6&     0.15&     0.37\\
  9&   0.7393&      1.7&     5.22&     4.80& $\pm~$   0.0016&   0.0040&      0.2&      0.5&     0.19&     0.52\\
 10&   0.7740&      5.1&     4.80&     9.04& $\pm~$   0.0005&   0.0013&      0.2&      0.5&     0.06&     0.17\\
 11&   0.8035&      6.7&     4.24&     8.16& $\pm~$   0.0005&   0.0013&      0.3&      0.7&     0.03&     0.08\\
 12&   0.8280&      7.2&     4.13&     8.22& $\pm~$   0.0005&   0.0015&      0.3&      0.7&     0.03&     0.09\\
 13&   0.8702&      2.4&     2.57&     5.17& $\pm~$   0.0015&   0.0048&      0.2&      0.4&     0.15&     0.45\\
 14 & 0.9149 & 10.2 & 4.18 & 11.12& $\pm~$ 0.0004  & 0.0011 &  0.1 &  0.4 &   0.04 &  0.10 \\
&\phn &\phn &\phn &\phn &$+~~   0.0006 $ & $   0.0015 $ & $      0.1 $ & $      0.4 $ & $     0.06 $ & $     0.17  $\\[-1ex]
 \raisebox{1.5ex}{ 15  } & \raisebox{1.5ex}{   0.9595} & \raisebox{1.5ex}{      6.5 }& \raisebox{1.5ex}{     2.67 }& \raisebox{1.5ex}{     9.80}&
$-~~   0.0006 $ & $   0.0015 $ & $      0.1 $ & $      0.4 $ & $     0.06 $ & $     0.16 $\\[0.5ex]
 16&   1.0236&      4.2&     0.31&     8.74& $\pm~$   0.0007&   0.0021&      0.1&      0.4&     0.08&     0.22\\
 17&   1.0595&      2.3&     0.06&     5.00& $\pm~$   0.0015&   0.0042&      0.1&      0.4&     0.15&     0.41\\
 18&   1.1297&      1.9&     4.55&     4.00& $\pm~$   0.0024&   0.0064&      0.2&      0.5&     0.28&     0.72\\
 19&   1.1418&      6.2&     1.65&     8.24& $\pm~$   0.0007&   0.0016&      0.2&      0.5&     0.08&     0.21\\
 20&   1.5301&      1.2&     4.16&     4.04& $\pm~$   0.0021&   0.0051&      0.1&      0.3&     0.22&     0.56\\
 21 & 1.7761&1.8 & 5.56 & 5.65 & $\pm~$ 0.0017 & 0.0047 &  0.1 & 0.4 & 0.19 & 0.46 \\
&\phn &\phn &\phn &\phn &$+~~   0.0009 $ & $   0.0024 $ & $      0.2 $ & $      0.5 $ & $     0.10 $ & $     0.35  $\\[-1ex]
 \raisebox{1.5ex}{ 22  } & \raisebox{1.5ex}{   1.8208} & \raisebox{1.5ex}{      4.8 }& \raisebox{1.5ex}{     4.63 }& \raisebox{1.5ex}{     8.97}&
$-~~   0.0009 $ & $   0.0024 $ & $      0.2 $ & $      0.5 $ & $     0.10 $ & $     0.30 $\\[0.5ex]
 23&   1.8441&      5.1&     4.67&     9.11& $\pm~$   0.0010&   0.0027&      0.2&      0.7&     0.09&     0.24\\
 24&   1.8685&      2.1&     4.81&     5.08& $\pm~$   0.0019&   0.0054&      0.2&      0.5&     0.18&     0.49\\
&\phn &\phn &\phn &\phn &$+~~   0.0007 $ & $   0.0019 $ & $      0.2 $ & $      0.4 $ & $     0.08 $ & $     0.23  $\\[-1ex]
 \raisebox{1.5ex}{ 25  } & \raisebox{1.5ex}{   1.9066} & \raisebox{1.5ex}{      4.3 }& \raisebox{1.5ex}{     3.26 }& \raisebox{1.5ex}{     8.53}&
$-~~   0.0007 $ & $   0.0019 $ & $      0.2 $ & $      0.4 $ & $     0.08 $ & $     0.19 $\\[0.5ex]
 26&   2.0171&      1.1&     3.44&     3.87& $\pm~$   0.0028&   0.0076&      0.1&      0.4&     0.32&     0.79\\
&\phn &\phn &\phn &\phn &$+~~   0.0006 $ & $   0.0015 $ & $      0.1 $ & $      0.4 $ & $     0.08 $ & $     0.29  $\\[-1ex]
 \raisebox{1.5ex}{ 27  } & \raisebox{1.5ex}{   2.0512} & \raisebox{1.5ex}{      6.1 }& \raisebox{1.5ex}{     1.91 }& \raisebox{1.5ex}{     9.38}&
$-~~   0.0006 $ & $   0.0015 $ & $      0.1 $ & $      0.4 $ & $     0.08 $ & $     0.19 $\\[0.5ex]
 28&   2.1160&      1.8&     0.11&     5.10& $\pm~$   0.0014&   0.0036&      0.1&      0.4&     0.15&     0.40\\
 29&   2.7814&      1.5&     2.77&     5.22& $\pm~$   0.0018&   0.0049&      0.1&      0.4&     0.21&     0.54\\
&\phn &\phn &\phn &\phn &$+~~   0.0029 $ & $   0.0072 $ & $      0.1 $ & $      0.3 $ & $     0.34 $ & $     1.01  $\\[-1ex]
 \raisebox{1.5ex}{ 30  } & \raisebox{1.5ex}{   2.8771} & \raisebox{1.5ex}{      1.0 }& \raisebox{1.5ex}{     6.18 }& \raisebox{1.5ex}{     3.55}&
$-~~   0.0029 $ & $   0.0086 $ & $      0.1 $ & $      0.3 $ & $     0.34 $ & $     1.01$
\enddata
\tablenotetext{a}{  
Phases are referenced to the first observation in the data set.}
\end{deluxetable}
\end{tiny}

\clearpage

\begin{tiny}
\begin{deluxetable}{c c c c  c cc cc cc}
\tablecolumns{11}
\tablewidth{0pc}
\tablecaption{HD 217543 periodicities from {\it MOST} photometry \label{tab_hd217543}}
\tablehead{
 \# &$\nu$ [c $\textrm{d}^{-1}$] &A [mmag] &$\phi$ [rad] &$\sim$ S/N &$\sigma_{\nu}$ &
 $3\sigma_{\nu}$ &$\sigma_{A}$ &$3\sigma_{A}$ &$\sigma_{\phi}$ &$3\sigma_{\phi}$
 }
\startdata
&\phn &\phn &\phn &\phn &$+~~   0.0003 $ & $   0.0035 $ & $      0.2 $ & $      1.2 $ & $     0.07 $ & $     0.58  $\\[-1ex]
 \raisebox{1.5ex}{  1  } & \raisebox{1.5ex}{   0.0269} & \raisebox{1.5ex}{     12.9 }& \raisebox{1.5ex}{     1.01 }& \raisebox{1.5ex}{    10.80}&
$-~~   0.0003 $ & $   0.0066 $ & $      0.2 $ & $      0.7 $ & $     0.07 $ & $     0.30 $\\[0.5ex]
&\phn &\phn &\phn &\phn &$+~~   0.0013 $ & $   0.0058 $ & $      0.3 $ & $      0.7 $ & $     0.14 $ & $     0.43  $\\[-1ex]
 \raisebox{1.5ex}{  2  } & \raisebox{1.5ex}{   0.0806} & \raisebox{1.5ex}{      7.1 }& \raisebox{1.5ex}{     0.09 }& \raisebox{1.5ex}{     5.48}&
$-~~   0.0013 $ & $   0.0031 $ & $      0.3 $ & $      0.7 $ & $     0.14 $ & $     0.43 $\\[0.5ex]
  3 & 0.1201 & 13.3 & 1.76 & 8.28& $\pm~$ 0.0012 &  0.0043 & 0.2  & 0.7  & 0.10 & 0.36\\
  4&   0.1806&      5.7&     3.03&     4.59& $\pm~$   0.0016&   0.0071&      0.3&      0.8&     0.12&     0.57\\
  5&   0.2454&      4.6&     3.97&     5.10& $\pm~$   0.0017&   0.0074&      0.3&      0.9&     0.12&     0.62\\
  6 &   0.2904&      1.9&     2.03&     3.15& $\pm~$   0.0042&   0.0121&      0.2&      1.7&     0.27&     1.01\\
  7 &   0.3133&      3.1&     4.67&     4.40& $\pm~$   0.0024&   0.0099&      0.3&      2.2&     0.19&     0.75\\
  8&   0.3987&      2.0&     5.78&     3.97& $\pm~$   0.0022&   0.0053&      0.2&      0.6&     0.19&     0.59\\
  9&   0.5147&      1.6&     2.80&     3.32& $\pm~$   0.0034&   0.0091&      0.2&      0.6&     0.31&     0.82\\
 10&   0.5668&      1.7&     4.37&     3.17& $\pm~$   0.0034&   0.0091&      0.2&      0.6&     0.32&     0.79\\
 11&   0.6154&      2.0&     5.61&     3.09& $\pm~$   0.0026&   0.0062&      0.2&      0.5&     0.23&     0.56\\
 12&   0.7772&      1.3&     1.87&     3.11& $\pm~$   0.0026&   0.0068&      0.2&      0.5&     0.23&     0.61\\
 13&   1.0006&      2.1&     2.68&     3.67& $\pm~$   0.0019&   0.0051&      0.2&      0.5&     0.16&     0.42\\
 14&   1.5826&      2.4&     0.72&     4.36& $\pm~$   0.0016&   0.0048&      0.2&      0.5&     0.16&     0.44\\
 15&   1.7164&      2.2&     1.85&     4.21& $\pm~$   0.0020&   0.0053&      0.2&      0.5&     0.19&     0.52\\
&\phn &\phn &\phn &\phn &$+~~   0.0019 $ & $   0.0054 $ & $      0.2 $ & $      0.5 $ & $     0.17 $ & $     0.69  $\\[-1ex]
 \raisebox{1.5ex}{ 16  } & \raisebox{1.5ex}{   1.7948} & \raisebox{1.5ex}{      2.6 }& \raisebox{1.5ex}{     0.45 }& \raisebox{1.5ex}{     4.16}&
$-~~   0.0019 $ & $   0.0074 $ & $      0.2 $ & $      0.5 $ & $     0.17 $ & $     0.46 $\\[0.5ex]
&\phn &\phn &\phn &\phn &$+~~   0.0011 $ & $   0.0038 $ & $      0.3 $ & $      1.6 $ & $     0.10 $ & $     0.55  $\\[-1ex]
 \raisebox{1.5ex}{ 17  } & \raisebox{1.5ex}{   1.8881} & \raisebox{1.5ex}{      6.0 }& \raisebox{1.5ex}{     1.81 }& \raisebox{1.5ex}{     5.66}&
$-~~   0.0011 $ & $   0.0038 $ & $      0.3 $ & $      0.8 $ & $     0.10 $ & $     0.44 $\\[0.5ex]
&\phn &\phn &\phn &\phn &$+~~   0.0006 $ & $   0.0014 $ & $      0.2 $ & $      0.7 $ & $     0.05 $ & $     0.48  $\\[-1ex]
 \raisebox{1.5ex}{ 18  } & \raisebox{1.5ex}{   1.9237} & \raisebox{1.5ex}{     12.7 }& \raisebox{1.5ex}{     3.51 }& \raisebox{1.5ex}{     8.17}&
$-~~   0.0006 $ & $   0.0056 $ & $      0.2 $ & $      0.7 $ & $     0.05 $ & $     0.18 $\\[0.5ex]
&\phn &\phn &\phn &\phn &$+~~   0.0009 $ & $   0.0209 $ & $      0.4 $ & $      3.6 $ & $     0.08 $ & $     0.19  $\\[-1ex]
 \raisebox{1.5ex}{ 19  } & \raisebox{1.5ex}{   1.9643} & \raisebox{1.5ex}{      4.9 }& \raisebox{1.5ex}{     4.44 }& \raisebox{1.5ex}{     5.07}&
$-~~   0.0009 $ & $   0.0027 $ & $      0.4 $ & $      1.8 $ & $     0.08 $ & $     1.28 $\\[0.5ex]
&\phn &\phn &\phn &\phn &$+~~   0.0008 $ & $   0.0029 $ & $      0.3 $ & $      4.7 $ & $     0.05 $ & $     0.17  $\\[-1ex]
 \raisebox{1.5ex}{ 20  } & \raisebox{1.5ex}{   1.9971} & \raisebox{1.5ex}{      7.6 }& \raisebox{1.5ex}{     6.22 }& \raisebox{1.5ex}{     6.80}&
$-~~   0.0008 $ & $   0.0029 $ & $      0.3 $ & $      0.8 $ & $     0.05 $ & $     0.41 $\\[0.5ex]
&\phn &\phn &\phn &\phn &$+~~   0.0002 $ & $   0.0007 $ & $      0.3 $ & $      0.9 $ & $     0.02 $ & $     0.39  $\\[-1ex]
 \raisebox{1.5ex}{ 21  } & \raisebox{1.5ex}{   2.0335} & \raisebox{1.5ex}{     23.8 }& \raisebox{1.5ex}{     2.25 }& \raisebox{1.5ex}{    14.25}&
$-~~   0.0002 $ & $   0.0036 $ & $      0.3 $ & $      5.3 $ & $     0.02 $ & $     0.02 $\\[0.5ex]
&\phn &\phn &\phn &\phn &$+~~   0.0010 $ & $   0.0031 $ & $      0.3 $ & $      3.9 $ & $     0.07 $ & $     1.18  $\\[-1ex]
 \raisebox{1.5ex}{ 22  } & \raisebox{1.5ex}{   2.0704} & \raisebox{1.5ex}{      7.1 }& \raisebox{1.5ex}{     5.30 }& \raisebox{1.5ex}{     6.26}&
$-~~   0.0010 $ & $   0.0102 $ & $      0.3 $ & $      0.7 $ & $     0.07 $ & $     0.28 $\\[0.5ex]
&\phn &\phn &\phn &\phn &$+~~   0.0006 $ & $   0.0066 $ & $      0.2 $ & $      1.1 $ & $     0.05 $ & $     0.13  $\\[-1ex]
 \raisebox{1.5ex}{ 23  } & \raisebox{1.5ex}{   2.1237} & \raisebox{1.5ex}{      9.9 }& \raisebox{1.5ex}{     2.50 }& \raisebox{1.5ex}{     7.28}&
$-~~   0.0006 $ & $   0.0012 $ & $      0.2 $ & $      0.7 $ & $     0.05 $ & $     0.52 $\\[0.5ex]
&\phn &\phn &\phn &\phn &$+~~   0.0006 $ & $   0.0081 $ & $      0.3 $ & $      0.7 $ & $     0.06 $ & $     0.38  $\\[-1ex]
 \raisebox{1.5ex}{ 24  } & \raisebox{1.5ex}{   2.1766} & \raisebox{1.5ex}{      9.4 }& \raisebox{1.5ex}{     4.20 }& \raisebox{1.5ex}{     6.80}&
$-~~   0.0006 $ & $   0.0038 $ & $      0.3 $ & $      0.9 $ & $     0.06 $ & $     0.53 $\\[0.5ex]

\tablebreak

&\phn &\phn &\phn &\phn &$+~~   0.0010 $ & $   0.0037 $ & $      0.2 $ & $      0.6 $ & $     0.08 $ & $     0.39  $\\[-1ex]
 \raisebox{1.5ex}{ 25  } & \raisebox{1.5ex}{   2.2288} & \raisebox{1.5ex}{      6.5 }& \raisebox{1.5ex}{     6.03 }& \raisebox{1.5ex}{     5.68}&
$-~~   0.0010 $ & $   0.0037 $ & $      0.2 $ & $      1.0 $ & $     0.08 $ & $     0.39 $\\[0.5ex]
&\phn &\phn &\phn &\phn &$+~~   0.0018 $ & $   0.0047 $ & $      0.2 $ & $      0.5 $ & $     0.18 $ & $     0.48  $\\[-1ex]
 \raisebox{1.5ex}{ 26  } & \raisebox{1.5ex}{   2.2985} & \raisebox{1.5ex}{      2.8 }& \raisebox{1.5ex}{     1.60 }& \raisebox{1.5ex}{     4.06}&
$-~~   0.0018 $ & $   0.0047 $ & $      0.2 $ & $      0.5 $ & $     0.18 $ & $     0.42 $\\[0.5ex]
 27&   2.3674&      3.0&     1.59&     4.27& $\pm~$   0.0016&   0.0046&      0.2&      0.6&     0.15&     0.40\\
 28&   3.7755&      2.0&     4.00&     4.00& $\pm~$   0.0019&   0.0049&      0.2&      0.5&     0.18&     0.51\\
&\phn &\phn &\phn &\phn &$+~~   0.0019 $ & $   0.0045 $ & $      0.2 $ & $      0.5 $ & $     0.18 $ & $     0.56  $\\[-1ex]
 \raisebox{1.5ex}{ 29  } & \raisebox{1.5ex}{   3.8947} & \raisebox{1.5ex}{      2.3 }& \raisebox{1.5ex}{     0.17 }& \raisebox{1.5ex}{     4.33}&
$-~~   0.0019 $ & $   0.0045 $ & $      0.2 $ & $      0.5 $ & $     0.18 $ & $     0.42 $\\[0.5ex]
 30&   4.0017&      3.4&     2.13&     4.52& $\pm~$   0.0014&   0.0037&      0.2&      0.5&     0.13&     0.35\\
&\phn &\phn &\phn &\phn &$+~~   0.0015 $ & $   0.0045 $ & $      0.2 $ & $      0.6 $ & $     0.15 $ & $     0.56  $\\[-1ex]
 \raisebox{1.5ex}{ 31  } & \raisebox{1.5ex}{   4.0646} & \raisebox{1.5ex}{      4.2 }& \raisebox{1.5ex}{     1.18 }& \raisebox{1.5ex}{     5.20}&
$-~~   0.0015 $ & $   0.0048 $ & $      0.2 $ & $      0.6 $ & $     0.15 $ & $     0.37 $\\[0.5ex]
 32&   4.1079&      2.6&     1.03&     4.08& $\pm~$   0.0025&   0.0068&      0.2&      0.6&     0.20&     0.52\\
&\phn &\phn &\phn &\phn &$+~~   0.0011 $ & $   0.0031 $ & $      0.2 $ & $      0.6 $ & $     0.12 $ & $     0.73  $\\[-1ex]
 \raisebox{1.5ex}{ 33  } & \raisebox{1.5ex}{   4.1886} & \raisebox{1.5ex}{      5.9 }& \raisebox{1.5ex}{     4.84 }& \raisebox{1.5ex}{     6.01}&
$-~~   0.0011 $ & $   0.0060 $ & $      0.2 $ & $      0.7 $ & $     0.12 $ & $     0.24 $\\[0.5ex]
&\phn &\phn &\phn &\phn &$+~~   0.0017 $ & $   0.0044 $ & $      0.2 $ & $      0.7 $ & $     0.16 $ & $     1.00  $\\[-1ex]
 \raisebox{1.5ex}{ 34  } & \raisebox{1.5ex}{   4.2350} & \raisebox{1.5ex}{      4.9 }& \raisebox{1.5ex}{     0.84 }& \raisebox{1.5ex}{     5.83}&
$-~~   0.0017 $ & $   0.0080 $ & $      0.2 $ & $      0.6 $ & $     0.16 $ & $     0.35 $\\[0.5ex]
&\phn &\phn &\phn &\phn &$+~~   0.0027 $ & $   0.0081 $ & $      0.2 $ & $      0.6 $ & $     0.27 $ & $     1.16  $\\[-1ex]
 \raisebox{1.5ex}{ 35  } & \raisebox{1.5ex}{   4.2665} & \raisebox{1.5ex}{      2.3 }& \raisebox{1.5ex}{     3.05 }& \raisebox{1.5ex}{     4.15}&
$-~~   0.0027 $ & $   0.0085 $ & $      0.2 $ & $      0.6 $ & $     0.27 $ & $     0.73 $\\[0.5ex]
&\phn &\phn &\phn &\phn &$+~~   0.0012 $ & $   0.0033 $ & $      0.2 $ & $      0.5 $ & $     0.14 $ & $     0.51  $\\[-1ex]
 \raisebox{1.5ex}{ 36  } & \raisebox{1.5ex}{   4.4017} & \raisebox{1.5ex}{      3.6 }& \raisebox{1.5ex}{     2.01 }& \raisebox{1.5ex}{     4.91}&
$-~~   0.0012 $ & $   0.0033 $ & $      0.2 $ & $      0.5 $ & $     0.14 $ & $     0.36 $\\[0.5ex]
&\phn &\phn &\phn &\phn &$+~~   0.0019 $ & $   0.0055 $ & $      0.2 $ & $      0.5 $ & $     0.20 $ & $     0.57  $\\[-1ex]
 \raisebox{1.5ex}{ 37  } & \raisebox{1.5ex}{   4.4695} & \raisebox{1.5ex}{      2.1 }& \raisebox{1.5ex}{     4.02 }& \raisebox{1.5ex}{     4.20}&
$-~~   0.0019 $ & $   0.0055 $ & $      0.2 $ & $      0.5 $ & $     0.20 $ & $     0.41 $\\[0.5ex]
&\phn &\phn &\phn &\phn &$+~~   0.0020 $ & $   0.0056 $ & $      0.2 $ & $      0.5 $ & $     0.18 $ & $     0.55  $\\[-1ex]
 \raisebox{1.5ex}{ 38  } & \raisebox{1.5ex}{   4.5406} & \raisebox{1.5ex}{      2.4 }& \raisebox{1.5ex}{     5.24 }& \raisebox{1.5ex}{     4.26}&
$-~~   0.0020 $ & $   0.0056 $ & $      0.2 $ & $      0.5 $ & $     0.18 $ & $     0.52 $\\[0.5ex]
 39&   4.6210&      1.6&     5.40&     3.38& $\pm~$   0.0029&   0.0072&      0.2&      0.5&     0.27&     0.73\\
&\phn &\phn &\phn &\phn &$+~~   0.0016 $ & $   0.0042 $ & $      0.2 $ & $      0.5 $ & $     0.17 $ & $     0.47  $\\[-1ex]
 \raisebox{1.5ex}{ 40  } & \raisebox{1.5ex}{   4.7226} & \raisebox{1.5ex}{      2.6 }& \raisebox{1.5ex}{     4.63 }& \raisebox{1.5ex}{     4.47}&
$-~~   0.0016 $ & $   0.0042 $ & $      0.2 $ & $      0.5 $ & $     0.17 $ & $     0.46$
\enddata
\end{deluxetable}
\end{tiny}

\end{document}